\documentclass[conference, a4paper]{IEEEtran}
\IEEEoverridecommandlockouts
\usepackage{cite}
\usepackage{amsmath,amssymb,amsfonts}
\usepackage{algorithmic}
\usepackage{graphicx}
\usepackage{textcomp}
\usepackage{xcolor}
\usepackage{gensymb} 
\usepackage{subfig} 
\usepackage{soul} 
\usepackage{hhline} 
\usepackage{main}
\usepackage{accents} 
\usepackage{xfrac} 
\usepackage{booktabs} 
\usepackage{caption}
\usepackage{xspace}
\usepackage{bm} 
\usepackage{multirow} 
\usepackage{hyperref}
\usepackage{transparent} 
\def\BibTeX{{\rm B\kern-.05em{\sc i\kern-.025em b}\kern-.08em
    T\kern-.1667em\lower.7ex\hbox{E}\kern-.125emX}}

\newcommand*\eg{\emph{e.g.}\xspace}
\newcommand*\ie{\emph{i.e.}\xspace}

\graphicspath{{./Figures/}}
\begin{document}

\title{Road Roughness Estimation via Fusion \\ of Standard Onboard Automotive Sensors
\thanks{
G. Zetterqvist has received funding from ELLIIT. 
This work was partially funded by the Wallenberg AI, Autonomous Systems and Software Program (WASP) funded by the Knut and Alice Wallenberg Foundation.
The authors would like to thank NIRA Dynamics AB for providing the data used in this study.
}
}

\author{Martin Agebjär$^\dagger$, Gustav Zetterqvist$^{\star}$, Fredrik Gustafsson$^\star$, Johan Wahlström$^\dagger$, Gustaf Hendeby$^\star$ \\
$^\star$ Dept. of Electrical Engineering, Linköping University, Linköping, Sweden \\
$^\dagger$ NIRA Dynamics AB, Linköping, Sweden \\
Email: \small \texttt{\{gustav.zetterqvist, fredrik.gustafsson, gustaf.hendeby\}@liu.se}\\
\texttt{\{martin.agebjar, johan.wahlstrom\}@niradynamics.se},
\\
}

\maketitle

\begin{abstract}
Road roughness significantly affects vehicle vibrations and ride quality. We introduce a \gls{kf}-based method for estimating road roughness in terms of the \gls{iri} by fusing inertial and speed measurements, offering a cost-effective solution for pavement monitoring.  
The method involves system identification on a physical vehicle to estimate realistic model parameters, followed by \gls{kf}-based reconstruction of the longitudinal road profile to compute \gls{iri} values. It explores \gls{iri} estimation using vertical and lateral vibrations, the latter more common in modern vehicles.
Validation on 230~km of real-world data shows promising results, with \gls{iri} estimation errors ranging from 1\% to 10\% of the reference values. However, accuracy deteriorates significantly when using only lateral vibrations, highlighting their limitations. These findings demonstrate the potential of \gls{kf}-based estimation for efficient road roughness monitoring.  
\end{abstract}


\pagenumbering{arabic}
\glsresetall

\section{Background} \label{sec:background}
Road roughness is the primary cause of vehicle vibrations, leading to reduced comfort, increased vehicle wear, and higher fuel consumption \cite{iri_fuel_consumption}. It is also strongly correlated with increased accident risk \cite{iri_safety}, accelerates pavement deterioration via dynamic loads \cite{iri_dynamic_load}, and disrupts vehicular communication links \cite{iri_V2V_communication}. Reliable information about road roughness is crucial for evidence-based decision-making by road operators. However, obtaining high-quality road measurements typically requires specialized vehicles and costly laser scanning systems. To address this challenge, NIRA Dynamics AB is developing an alternative cost-effective pavement monitoring solution that enables large-scale roughness data collection.

\begin{figure}[t]
    \centering
    \includegraphics[width=0.95\columnwidth]{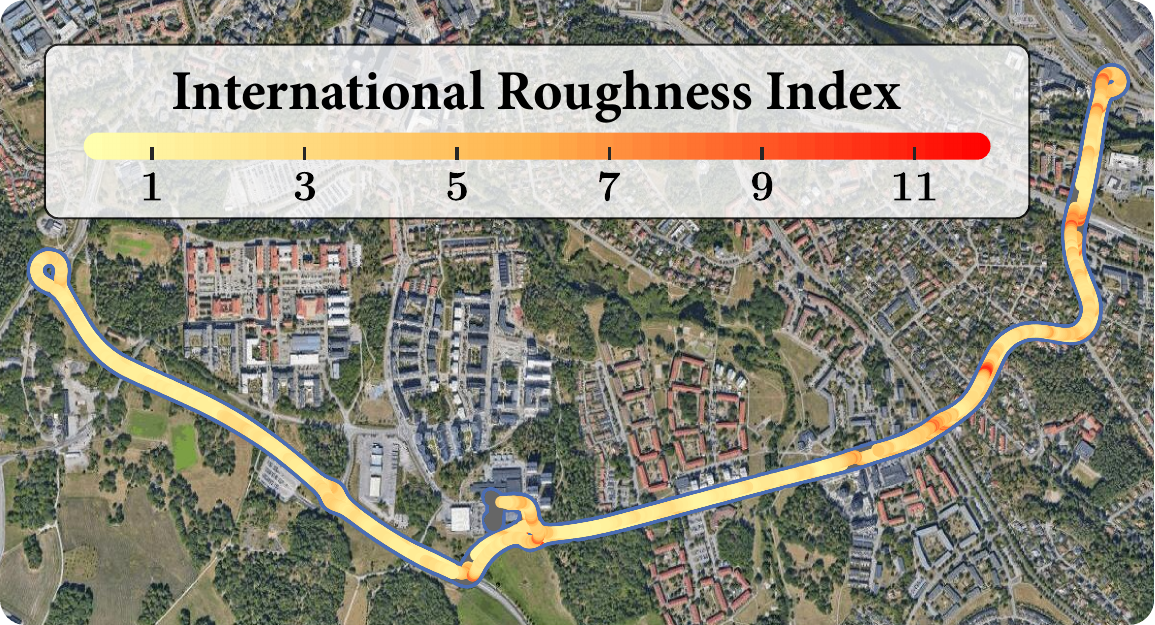}
    \caption{Example of a road profile and the corresponding \gls{iri} values for a road segment in Linköping, Sweden.
    Higher \gls{iri} values represent rougher stretches of the road.}    
    \label{fig:IRI_example}
    \vspace{-0.5cm}
\end{figure}

A widely used measure of road roughness is the \gls{iri}, which quantifies road surface irregularities based on a given longitudinal road profile \cite{iri}. The \gls{iri} was specifically designed to correlate with vertical passenger acceleration, representing the vibrations experienced inside the vehicle \cite{little_book}. A cost-effective approach to estimating the \gls{iri} involves measuring these vibrations using the commonly available \gls{imu} in vehicles. 
The resulting road condition data, geotagged via the vehicle's built-in \gls{gnss} receiver, can contribute to large-scale cloud-based data analysis for road monitoring, as illustrated in \Figref{fig:IRI_example}.

The idea of using existing onboard sensors in modern cars to obtain virtual sensors with fusion software has been proposed for many purposes before, \eg, to fuse data for position \cite{Berntorp2022,Hall_2001}, road mapping \cite{Berntorp2023}, friction \cite{Gustafsson1997} and indirect tire pressure monitoring systems \cite{Persson2002}. 

Previous studies on road roughness estimation using vehicle vibrations rely primarily on vertical vibrations and often incorporate additional sensors, such as unsprung mass acceleration measurements, \ie, accelerations of components not supported by the suspension, like wheels and axles.
For example, \cite{one_phone} uses a \gls{qc} model to estimate the road profile but lacks clarity on how the sprung mass acceleration response is utilized. Similarly, \cite{several_phones} employs a longitudinal \gls{hc} model with multiple sensors, but provides limited detail on the estimation process. Other studies suggest a linear correlation between \gls{iri} and the \gls{rms} value of the relative vertical velocity between the sprung and unsprung masses \cite{est_and_speed_comp, correlation_proof}, using transfer functions and Parseval's theorem for estimation. Data-driven approaches have also been explored to refine transfer function gains \cite{interesting_TF_funtion}. In addition, observers and filters \cite{kalman_1, kalman_2, H_observer} have been used to estimate the road profile, but these methods require additional sensors. This makes the methods impractical for widespread vehicle-based monitoring, as the objective is to avoid using extra sensors.


This paper is based on a MSc thesis \cite{Agebjar_2024}, where a \gls{qc} and lateral \gls{hc} model are used to estimate the \gls{iri} from vertical and lateral vibrations, respectively. Its novelty lies in the complete analytical derivation from vehicle vibrations to road profile and \gls{iri} estimation. 
Additionally, to the best of our knowledge, this is the first study to estimate the road profile and \gls{iri} solely using lateral vibrations, given the car dynamics. These vibrations are more readily available on the vehicle's \gls{can} bus, as they are utilized by the \gls{esc} system \cite{ESC_Ford}, which is a significant advantage for practical implementation of data collection.

\section{Dynamic Models}
The dynamic models used to describe the relationship between the road profile and the measurements obtained from the vehicle's \gls{imu} are presented below.

\subsection{Quarter-Car Model} \label{sec:theory_QC_model}
\begin{figure}[tb]
\centering
\subfloat[\textit{Quarter-car} model \label{fig:QC_model}]{\includegraphics[width=0.29\columnwidth]{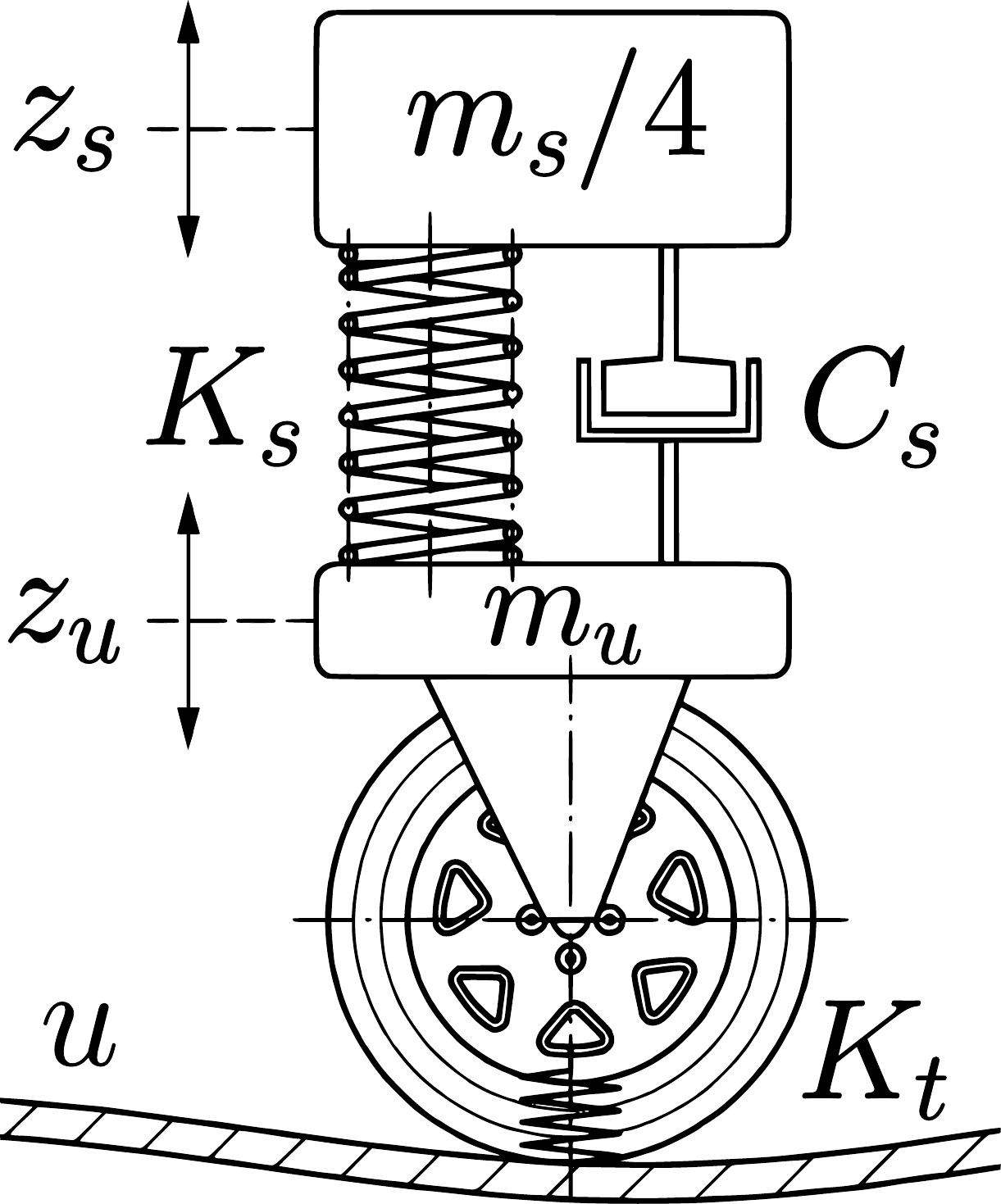}}
\hfill
\subfloat[Lateral \textit{half-car} model \label{fig:hc_model}]{\includegraphics[width=0.66\columnwidth]{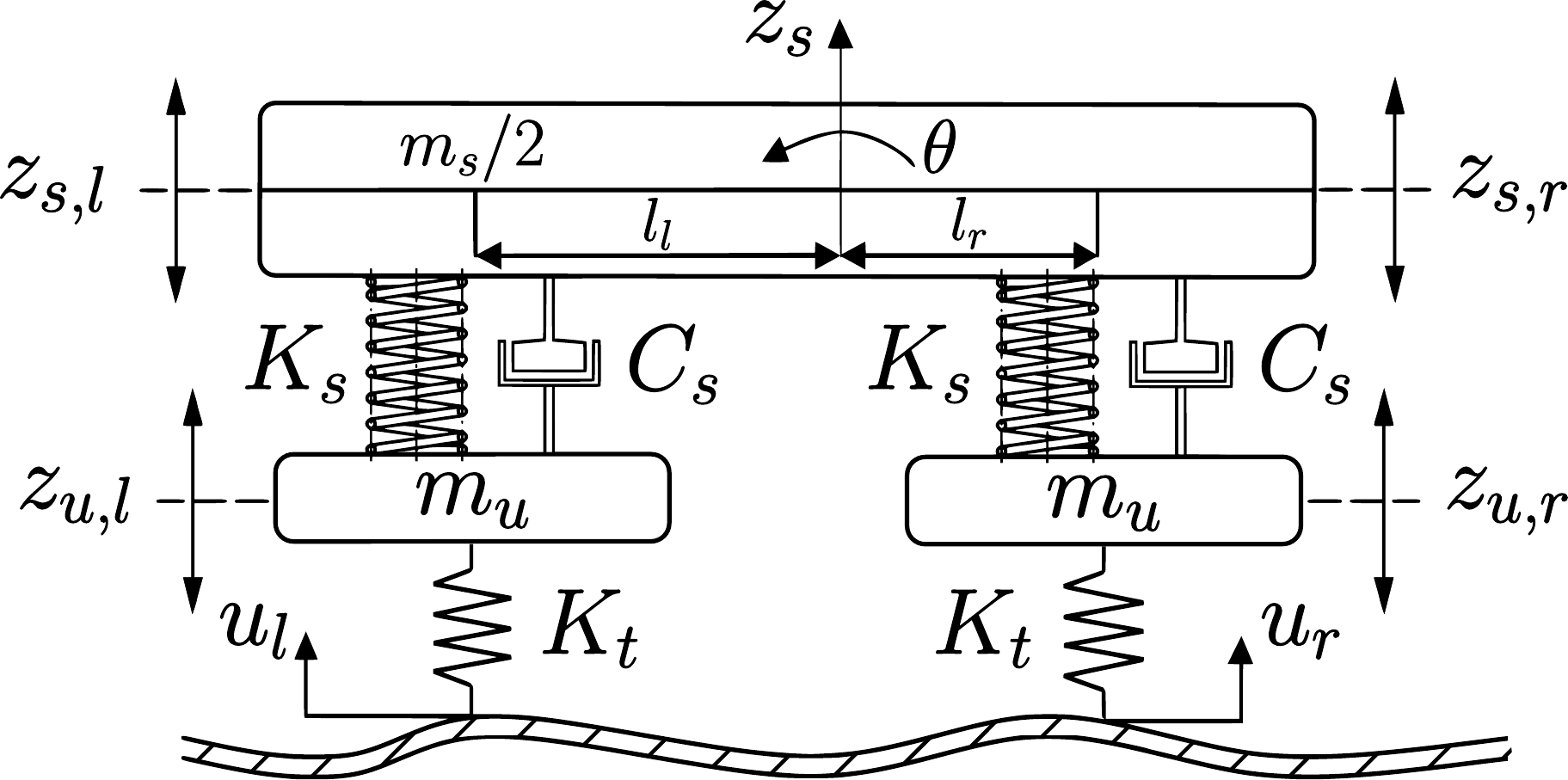}}
\caption{Dynamic car models used in this study. The \glsxtrshort{qc} model represents one-quarter of the vehicle, while the lateral \glsxtrshort{hc} model extends this representation to half of the vehicle. Both models share the same parameters: sprung mass ($m_s$), unsprung mass ($m_u$), suspension stiffness ($K_s$), suspension damping ($C_s$), and tire stiffness ($K_t$). Figure (a) and (b) is inspired by \cite{ss_standard} and \cite{hc_model}, respectively.}
\label{fig:dynamic_models}
\vspace{-0.1cm}
\end{figure}

The \glsxtrfull{qc} model is a widely used representation of a vehicle's vertical dynamics. The model and its associated parameters are illustrated in \Figref{fig:QC_model} \cite{ss_standard}. 
The differential equations describing the model's dynamics are,
\begin{align*}
\ddot{z}_s(t) m_s/4 &= - K_s(z_s(t) - z_u(t)) - C_s(\dot{z}_s(t) - \dot{z}_u(t)),  \\
\ddot{z}_u(t) m_u &= + K_s(z_s(t) - z_u(t)) + C_s(\dot{z}_s(t) - \dot{z}_u(t)) \\
&\qquad - K_t(z_u(t) - u(t)) ,
\end{align*}
where $z_s(t)$ is the sprung mass vertical displacement, $z_u(t)$ is the unsprung mass vertical displacement, $u(t)$ is the road elevation profile, and the dot denotes the time derivative. 
The parameters are the same as in \Figref{fig:QC_model}, where the
sprung mass is denoted by $m_s$, the unsprung mass by $m_u$, the suspension stiffness by $K_s$, the suspension damping by $C_s$, and the tire stiffness by $K_t$.
By introducing the state $\bm{x}_{QC}(t) =
\begin{bmatrix}
z_s(t) & \dot{z}_s(t) & z_u(t) & \dot{z}_u(t)
\end{bmatrix}^T$, the model's state-space representation is obtained,

\begin{equation*}
{\dot{\bm{x}}}_{QC}(t) =
\underbrace{
    \left[
\begin{smallmatrix}
    0 & 1 & 0 & 0 \\
    -\frac{4 K_s}{m_s} & -\frac{4 C_s}{m_s} & \frac{4 K_s}{m_s} & \frac{4 C_s}{m_s} \\
    0 & 0 & 0 & 1 \\
    \frac{K_s}{m_u} & \frac{C_s}{m_u} & -\frac{K_s+K_t}{m_u} & -\frac{C_s}{m_u}
\end{smallmatrix}\right]}_{\bm{A}_{QC}}
\bm{x}_{QC}(t)
+
\underbrace{
    \left[ 
\begin{smallmatrix}
    0 \\
    0 \\
    0 \\
    \frac{K_t}{m_u}
\end{smallmatrix}\right]}_{\bm{B}_{QC}}
u(t).
\end{equation*}

The measured vertical acceleration $y_{QC}(t)$ of the sprung mass is obtained as,
\begin{equation*}
\label{eq:QC_sys_out}
y_{QC}(t) = \ddot{z}_s(t) = 
\underbrace{
\begin{bmatrix}
    -\frac{4 K_s}{m_s} & -\frac{4 C_s}{m_s} & \frac{4 K_s}{m_s} & \frac{4 C_s}{m_s}
\end{bmatrix}}_{\bm{C}_{QC}}
\bm{x}_{QC}(t).
\end{equation*}

\subsection{Half-Car Model} \label{sec:theory_HC_model}
The \glsxtrfull{hc} model provides a more comprehensive representation of a vehicle's dynamics compared to the \gls{qc} model. The model and its associated parameters are illustrated in \Figref{fig:hc_model} \cite{hc_model}. The model is here used in a lateral configuration, representing a single vehicle axle, incorporating one left and one right wheel, which allows it to capture how differences between the left and the right wheel tracks affect roll movements $\theta(t)$ and vibrations of its sprung mass. Note that commonly denoted parameters are equivalent in the \gls{qc} and \gls{hc} model.

The differential equations describing the model's dynamics, assuming small angles for $\theta$, are,
\begin{align*}
    \ddot{z}_s m_s/2 &= -K_s(z_{s,l} - z_{u,l}) - K_s(z_{s,r} - z_{u,r}) \\
    &\qquad - C_s(\dot{z}_{s,l} - \dot{z}_{u,l}) - C_s(\dot{z}_{s,r} - \dot{z}_{u,r}), \\
    I_{s} \ddot{\theta} &= l_l (K_s(z_{s,l} - z_{u,l}) + C_s(\dot{z}_{s,l} - \dot{z}_{u,l})) \\
    &\qquad - l_r (K_s(z_{s,r} - z_{u,r}) + C_s(\dot{z}_{s,r} - \dot{z}_{u,r})), \\
    \ddot{z}_{u,l} m_{u} &= K_s(z_{s,l} - z_{u,l}) + C_s(\dot{z}_{s,l} - \dot{z}_{u,l}) - K_t(z_{u,l} - u_l), \\
    \ddot{z}_{u,r} m_{u} &= K_s(z_{s,r} - z_{u,r}) + C_s(\dot{z}_{s,r} - \dot{z}_{u,r}) - K_t(z_{u,r} - u_r), 
\end{align*}
where $I_s$ is the sprung mass' moment of inertia, $z_{s,l}(t)$ and $z_{s,r}(t)$ denote the left and right vertical displacement of sprung mass, respectively, $z_{u,l}(t)$ and $z_{u,r}(t)$ refer to the left and right vertical displacement of unsprung mass, respectively, and $u_l(t)$ and $u_r(t)$ correspond to the left and right road elevation profile, respectively. 
Additionally, $l_l$ and $l_r$ represent the distances from the left and right wheel to the \gls{imu}, respectively, as depicted in \Figref{fig:hc_model}.
The vertical displacement of the sprung mass $z_s(t)$ and the roll angle $\theta(t)$ are given by
\begin{equation*}
    z_s(t) = \frac{l_l z_{s,r}(t) + l_r z_{s,l}(t)}{l_l + l_r}, \quad \theta(t) = \frac{z_{s,r}(t) - z_{s,l}(t)}{l_l + l_r}.
\end{equation*}

If the \gls{imu} is placed centered inside the vehicle, \ie, $l_l = l_r = l$, the equations can be simplified and here expressed in terms of $z_s(t)$ and $\theta(t)$ instead of $z_{s,l}(t)$ and $z_{s,r}(t)$,
\begin{align*}
    \ddot{z}_s m_s/2 &= -K_s(2 z_s - z_{u,l} - z_{u,r}) - C_s(2 \dot{z}_s - \dot{z}_{u,l} - \dot{z}_{u,r}), \\
    I_{s} \ddot{\theta} &= K_s l(z_{u,r} - z_{u,l} - 2 l \theta) + C_s l(\dot{z}_{u,r} - \dot{z}_{u,l} - 2 l \dot{\theta}), \\
    \ddot{z}_{u,l} m_{u} &= K_s(z_s - z_{u,l} - l \theta) + C_s(\dot{z}_s - \dot{z}_{u,l} - l \dot{\theta}) \\
    &\qquad  - K_t(z_{u,l} - u_l), \\
    \ddot{z}_{u,r} m_{u} &= K_s(z_s - z_{u,r} + l \theta) + C_s(\dot{z}_s - \dot{z}_{u,r} + l \dot{\theta}) \\
    &\qquad  - K_t(z_{u,r} - u_r).
\end{align*}

By introducing $$\setlength{\arraycolsep}{4pt} \bm{x}_{HC}(t) = \left[\begin{smallmatrix}
z_s(t) & \dot{z}_s(t) & \theta(t) & \dot{\theta}(t) & z_{u,l}(t) & \dot{z}_{u,l}(t) & z_{u,r}(t) & \dot{z}_{u,r}(t)
\end{smallmatrix}\right]^T,$$ the model's state-space representation is obtained,
\begin{equation}
\label{eq:HC_sys}
\dot{\bm{x}}_{HC}(t) = 
\bm{A}_{HC}
\bm{x}_{HC}(t) +\bm{B}_{HC}
\bm{u}(t), \\
\end{equation}
\begin{align*}
    {\scriptstyle\bm{A}_{HC}} & \scriptstyle{=} 
\begin{bmatrix}
    \begin{smallmatrix}
    0 & 1 & 0 & 0 & 0 & 0 & 0 & 0 \\
    \frac{-4 K_s}{m_s} & \frac{-4 C_s}{m_s} & 0 & 0 & \frac{2 K_s}{m_s} & \frac{2 C_s}{m_s} & \frac{2 K_s}{m_s} & \frac{2 C_s}{m_s} \\
    0 & 0 & 0 & 1 & 0 & 0 & 0 & 0 \\
    0 & 0 & \frac{-2 K_s l^2}{I_s} & \frac{-2 C_s l^2}{I_s} & \frac{- K_s l}{I_s} & \frac{- C_s l}{I_s} & \frac{K_s l}{I_s} & \frac{C_s l}{I_s} \\
    0 & 0 & 0 & 0 & 0 & 1 & 0 & 0 \\
    \frac{K_s}{m_u} & \frac{C_s}{m_u} & \frac{-K_s l}{m_u} & \frac{-C_s l}{m_u} & \frac{-K_s - K_t}{m_u} & \frac{-C_s}{m_u} & 0 & 0 \\
    0 & 0 & 0 & 0 & 0 & 0 & 0 & 1 \\
    \frac{K_s}{m_u} & \frac{C_s}{m_u} & \frac{K_s l}{m_u} & \frac{C_s l}{m_u} & 0 & 0 & \frac{-K_s - K_t}{m_u} & \frac{-C_s}{m_u}
    \end{smallmatrix}
\end{bmatrix},
\\
{\scriptstyle\bm{B}_{HC}} & \scriptstyle{=} 
\begin{bmatrix}
    \begin{smallmatrix}
0 & 0 & 0 & 0 & 0 & \frac{K_t}{m_u} & 0 & 0 \\
0 & 0 & 0 & 0 & 0 & 0 & 0 & \frac{K_t}{m_u}
\end{smallmatrix}\end{bmatrix}
^T ,
\end{align*}
where \(\bm{u}(t) =
\left[
\begin{smallmatrix}
u_l(t) & u_r(t)
\end{smallmatrix}\right]^T\) represents the left and right road profiles, which serve as inputs to the system.

The obtained measurements consist of vertical acceleration $\ddot{z}_s(t)$ and/or lateral acceleration $\ddot{x}_s(t)$ of the vehicle's sprung mass. The measured lateral acceleration is proportional to the roll angle acceleration $\ddot{\theta}(t)$, assuming small angles of $\theta(t)$. If the proportional constant is incorporated into the unknown moment of inertia $I_s$, the measurements $\bm{y}_{HC}(t)$ are represented as,
\begin{align*}
\bm{y}_{HC}(t)
&=
\begin{bmatrix}
\ddot{z}_s(t) \\
\ddot{x}_s(t)
\end{bmatrix} = {\bm{C}_{HC}}
\bm{x}_{HC}(t),
\\
\bm{C}_{HC}
&=
\begin{bmatrix}
\label{eq:HC_sys_out}
\begin{smallmatrix}
\frac{- 4 K_s}{m_s} & \frac{- 4 C_s}{m_s} & 0 & 0 & \frac{2 K_s}{m_s} & \frac{2 C_s}{m_s} & \frac{2 K_s}{m_s} & \frac{2 C_s}{m_s} \\
0 & 0 & \frac{- 2 K_s l^2}{I_s} & \frac{- 2 C_s l^2}{I_s} & \frac{- K_s l}{I_s} & \frac{- C_s l}{I_s} & \frac{K_s l}{I_s} & \frac{C_s l}{I_s}
\end{smallmatrix}
\end{bmatrix}.
\end{align*}
The measured vertical and lateral acceleration are rather perceived as vibrations and will hereafter be referred to as vertical and lateral vibration.

\section{International Roughness Index (IRI)} \label{sec:theory_iri}
The \glsxtrfull{iri} was established in 1986 and is based on previous work by the U.S. National Cooperative Highway Research Program \cite{iri}. 
The \gls{iri} quantifies the longitudinal roughness of a road profile, averaged over a specified distance 
$L$; for example, 40 meters.
The \gls{iri} is based on a certain virtual variant of the \gls{qc} model seen in \Figref{fig:QC_model}, often referred to as the ``Golden car.'' It is defined as the average absolute value of the derivative of the rattle space, $z_s(t) - z_u(t)$, over the time interval $L/V$, at a constant vehicle speed of $V = 80~\text{km/h}$,
\begin{equation}
\label{eq:iri}
  IRI = \frac{1}{L} \int_{0}^{L/V} \lvert \underbrace{\dot{z}_s(t) - \dot{z}_u(t)}_{\textstyle \xi(t)} \rvert \, dt,
\end{equation}
where $\xi(t)$ represents the derivative of the rattle space, $\dot{z}_s(t) - \dot{z}_u(t)$, as used in this paper.

The \gls{iri} is expressed in units of slope, $\text{mm/m}$ or $\text{m/km}$, which typically means that $\xi(t)$ has unit $\text{mm/s}$, $L$ has unit m, and $V$ has unit $\text{m/s}$. The specific ``Golden car'' model used for calculation of the \gls{iri} has certain parameters listed in Table~\ref{tab:golden_car}. The name was intended to convey this computer representation as a calibration reference \cite{little_book}. The \gls{iri} is defined as a property of a single wheel-track profile. When two profiles are measured at the same time, an alternative analysis is sometimes done by averaging the profiles, sample by sample, and then proceeding using the definition of the \gls{iri} \cite{iri}. 
\begin{table}[tb]
  \centering
  \caption{Vehicle dynamic parameters for the ``Golden car.''}
  \label{tab:golden_car}
  \begin{tabular}{lccc}
    \toprule
    Property & Value & Unit \\
    \midrule
    Total Sprung Mass ($m_s$) & 1\,000 & kg \\
    Unsprung Mass ($m_u$) & 37.5 & kg \\
    Suspension Spring Stiffness ($K_s$) & 15\,825 & N/m \\
    Suspension Damping Coefficient ($C_s$) & 1\,500 & Ns/m \\
    Tire Stiffness ($K_t$) & 163\,250 & N/m \\
    \bottomrule
  \end{tabular}
\end{table}

\begin{figure}[tb]
    \centering
    \setlength\figurewidth{0.6\columnwidth}
    \setlength\figureheight{0.3\columnwidth}
%
%
\definecolor{mycolor1}{rgb}{0.00000,0.44700,0.74100}%
\begin{tikzpicture}

\begin{axis}[%
width=0.9511\figurewidth,
height=\figureheight,
at={(0\figurewidth,0\figureheight)},
scale only axis,
xmode=log,
xmin=0.1,
xmax=100,
xminorticks=true,
xlabel style={font=\small},
xlabel={Frequency [Hz]},
ymin=0,
ymax=5.5,
ylabel style={font=\small},
ylabel={Magnitude},
tick label style={font=\small},
xmajorgrids,
xminorgrids,
ymajorgrids,
grid style={dashed},
title={The \gls{iri} frequency gain}
]
\addplot [color=mycolor1, line width=1.5pt, forget plot]
  table[row sep=crcr]{%
0.099310918137498	0.000173600437420873\\
0.196468664618044	0.00136428085239793\\
0.266333272517498	0.00345626414436406\\
0.326222200971167	0.00647033143024611\\
0.378074666359935	0.0102669255680503\\
0.426215882901533	0.0150122477222432\\
0.471708469091701	0.0207936299331886\\
0.512518692705333	0.027241851115904\\
0.551749237612913	0.034743857707439\\
0.588531577519145	0.0431036051256291\\
0.622004882563471	0.051970299489958\\
0.657382014340959	0.0628017320232876\\
0.68839520696455	0.0736653036639821\\
0.720871503378214	0.0865505549989747\\
0.754879928165343	0.101852157901765\\
0.783238259917919	0.116141262368588\\
0.812661920009195	0.132537747337438\\
0.843190929286625	0.151330147338742\\
0.874866812047991	0.172817647462487\\
0.907732652521022	0.197289027332834\\
0.941833153464794	0.224987522529486\\
0.977214696972572	0.256057582612252\\
1.01392540755881	0.290472207266522\\
1.061759183483	0.337728794618314\\
1.13254131515281	0.409205050265793\\
1.21923125164911	0.490688417902335\\
1.27675070431926	0.536600137908629\\
1.32471398786612	0.568781705134797\\
1.37447909267754	0.596442775976853\\
1.42611370719413	0.619723724974488\\
1.4796880626864	0.639175856169493\\
1.53527502878042	0.655576419405935\\
1.60770442167382	0.673057483961886\\
1.71488196987054	0.694522202425602\\
1.91550055557353	0.730919000991036\\
2.0244465099768	0.751358651915035\\
2.13958887134342	0.774211247826059\\
2.24052786930002	0.795408789830363\\
2.34622884814226	0.818767477074536\\
2.45691646298279	0.844448904933082\\
2.57282596744793	0.872604163365841\\
2.69420371368188	0.903385452502271\\
2.82130767593947	0.936954743103221\\
2.95440799888038	0.973490641352611\\
3.09378757173014	1.01319428611035\\
3.2397426295282	1.05629487835114\\
3.39258338274099	1.10305529370734\\
3.55263467657814	1.15377813780793\\
3.72023668141307	1.2088125528464\\
3.8957456157755	1.26856205878869\\
4.07953450345245	1.33349370216813\\
4.27199396630678	1.40414877732815\\
4.47353305449846	1.481155363092\\
4.68458011587305	1.56524285670142\\
4.90558370636506	1.65725854305242\\
5.13701354335134	1.75818593460619\\
5.3793615039807	1.86916402147254\\
5.63314267060136	1.99150544569498\\
5.89889642550851	2.12670954694863\\
6.17718759733849	2.27646251362619\\
6.46860766154632	2.44261034289522\\
6.71161176749627	2.58862834139961\\
7.02824426430835	2.7890858002939\\
7.35981447526576	3.01086538050207\\
7.7070271142123	3.25457259764222\\
8.07062014114951	3.51895405681419\\
8.60864769614924	3.91411574206993\\
9.18254283565628	4.30839450436048\\
9.52750047242728	4.51103449341053\\
9.79469667069539	4.64084270774633\\
9.97697764236321	4.71299661783646\\
10.16265089393	4.77137329409336\\
10.3517795563018	4.81428341555124\\
10.447659715608	4.8295070900676\\
10.5444279352617	4.84040188884082\\
10.6420924406472	4.8468766198105\\
10.7406615333343	4.84887444828607\\
10.8401435917833	4.84637417562217\\
10.9405470720574	4.83939073187894\\
11.0418805085416	4.82797486060775\\
11.1441525146679	4.81221200913961\\
11.35154708921	4.76814886638425\\
11.5628013120738	4.70849255504015\\
11.7779870119712	4.63491111497288\\
12.1082975023204	4.50267029405422\\
12.4478714618791	4.35034812865273\\
13.0351224468151	4.06978506963245\\
14.5600599502065	3.38164615796681\\
15.2469572701758	3.12012628573612\\
15.9662602210143	2.8804251562056\\
16.7194975973199	2.66289708024398\\
17.5082703173573	2.46643917602235\\
18.3342548256229	2.289285584476\\
19.1992066559329	2.1294596718021\\
20.104964162605	1.98501263139157\\
21.053452427667	1.85413729652377\\
22.0466873523941	1.73521267392941\\
23.0867799418717	1.62681194973199\\
24.1759407916913	1.5276925538086\\
25.3164847863136	1.43677848881107\\
26.5108360190854	1.35314032725762\\
27.761532944368	1.27597559108751\\
29.0712337727258	1.20459075546087\\
30.72468842709	1.12571865498747\\
32.4721849207313	1.05341982751754\\
34.3190719745904	0.986940779224813\\
36.2710025233065	0.925640074740801\\
38.333951017666	0.868968986069163\\
40.5142317111465	0.816455725826601\\
42.8185179865242	0.767692597603766\\
45.2538627817017	0.722325506969151\\
48.2707096560319	0.673279277233545\\
51.4886745013749	0.628009974926163\\
54.9211648388779	0.586149391013827\\
58.5824820015254	0.547377514347053\\
62.487880720069	0.511414549516085\\
66.6536326812492	0.478014458165338\\
71.7556091893694	0.442703224075276\\
77.248114514034	0.410169690799622\\
83.1610415323096	0.380162024538643\\
89.526571259964	0.352457496866775\\
96.379347996158	0.326858063366584\\
100	0.314791717689099\\
};
\end{axis}
\end{tikzpicture}%
    \caption{Gain from road elevation profile, the \gls{iri} effectively captures elevation frequencies between 3 and 33~Hz}
    \label{fig:iri_freq_resp}
    \vspace{-0.1cm}
\end{figure}
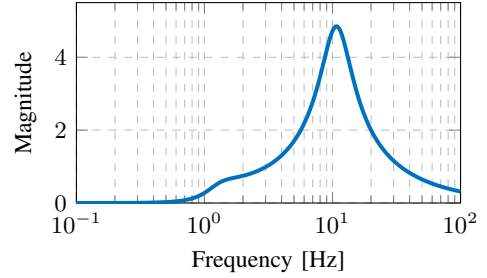

The ``Golden car'' system has the frequency responses shown in \Figref{fig:iri_freq_resp}. The index effectively captures roughness between 3 and 33~Hz. Consequently, roughness such as potholes, road patching, and other surface irregularities, impacts \gls{iri}, while a smooth hill corresponds to low frequencies that do not increase the index. An \gls{iri} value of zero corresponds to a perfectly flat road surface. While there is no theoretical upper limit for the \gls{iri}, a value around 10 to 12 mm/m indicates very high roughness.
\section{Experimental Setup}
Throughout the paper, two categories of data are used. The first category includes road-related data, such as laser-scanned road profile measurements obtained from a specialized vehicle, referred to as the ``laser car.'' These road profile measurements are captured at a spatial resolution of $S = 0.1$ meters and are linked to the ``laser car's'' current position via an onboard \gls{gnss} receiver.

The second category consists of data collected from an Audi A6 e-tron, referred to as the ``\gls{imu} car.'' This data includes vibration measurements obtained from the car’s \gls{imu}, as well as speed data. Measurements from the ``\gls{imu} car'' are collected at a sample rate of 50 Hz from the \gls{can} bus and are linked to the car's current position via its onboard \gls{gnss} receiver.

Both cars are driven over the same road stretches on different days, each attempting to follow the nominal position on the road. A total of 230.8~km of road data is collected in and around Linköping, Sweden, covering diverse road types, including highways, urban roads, and rural roads. The road data is divided into four \gls{iri} level categories, and the traversed distance in each \gls{iri} level category is shown later in \Tabref{tab:field_test_results}.

\subsection{GNSS Matching} \label{sec:method_GPS_matching}
\begin{figure}[tb]
    \centering
    \hspace{-0.3cm}
    \subfloat[\gls{gnss} matching overview.]{
        \setlength\figurewidth{0.8\columnwidth}
        \setlength\figureheight{0.5\columnwidth}
        \resizebox{0.53\columnwidth}{!}{\input{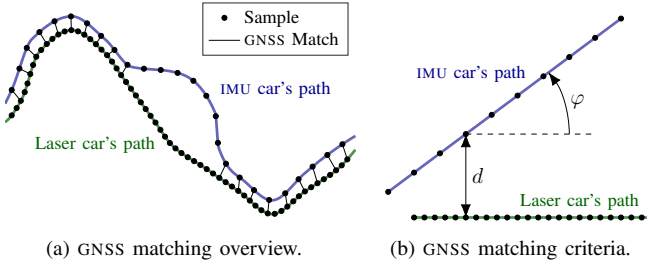}}
    }
    \hspace{-0.3cm}
    \subfloat[\gls{gnss} matching criteria.]{
        \setlength\figurewidth{0.8\columnwidth}
        \setlength\figureheight{0.5\columnwidth}
        \resizebox{0.4\columnwidth}{!}{\begin{tikzpicture}
  \draw[color={rgb,255:red,0; green,100; blue,0}, line width=1.5pt, opacity=0.7] (0.5,0) -- (5,0);
  \foreach \x in {0.5,0.7,...,5}
    \filldraw (\x,0) circle (1.3693pt);

  \draw[color={rgb,255:red,0; green,0; blue,139}, line width=1.5pt, opacity=0.6] (0,0.5) -- (4.5,3.875);
  \foreach \x in {0,0.5,...,4.5}
    \filldraw (\x,{\x*0.75 + 0.5}) circle (1.3694pt);

  \draw[dashed] (1.5, {1.5*0.75 + 0.5}) -- (4, {1.5*0.75 + 0.5});

\draw[-{Latex[scale=1.5]}]
  (3.5, {1.5*0.75 + 0.5}) arc[start angle=0, end angle=36.87, radius=2cm]
  node[midway, right] {\large  $\varphi$};  

\draw[{Latex[scale=1.5]}-{Latex[scale=1.5]}]
  (1.5,0) -- (1.5,{1.5*0.75 + 0.5})
  node[midway, right] {\large $d$};  

\node[text={rgb,255:red,0; green,0; blue,139}] at (1.6,2.8) {\gls{imu} car's path};

\node[text={rgb,255:red,0; green,100; blue,0}] at (3.7,0.3) {Laser car's path}; 

  \node at (current bounding box.south) [below, inner sep=2pt] {};
\end{tikzpicture}}
    }
    \caption{The \gls{gnss} matching algorithm links every \gls{imu} measurement to a road profile measurement if the Euclidean distance $d$ between their associated positions is not too large and the angle $\varphi$ between the ``laser car's'' and the \gls{imu} car's paths is not too large.}
    \label{fig:GPS_matching}
    \vspace{-0.1cm}
\end{figure}

\gls{gnss} matching is used to synchronize the data and ensure compatibility between measurements from the two categories. For each measurement sample in the ``\gls{imu} car'', the \gls{gnss} matching algorithm identifies the closest road profile measurement. A match occurs if the Euclidean distance between the positions of the two measurements is less than $d$ meters, and the angle between the paths of the ``laser car'' and the ``\gls{imu} car'' is less than $\varphi$ degrees. No match occurs if, for example, the vehicles are traveling in opposite directions on the same road. The \gls{gnss} matching process is illustrated in \Figref{fig:GPS_matching}. Although the exact accuracy of the specific \gls{gnss} receivers used is not fully known, an accuracy of at least four meters is expected, based on the general performance specifications of open service \gls{gnss} \cite{GNSS_presicion}.
\section{System Identification} \label{sec:method_system_identification}
To estimate the road profile, the model parameters of the ``\gls{imu} car'' must be identified through system identification.

The system identification method uses Grey-Box modeling, which combines a partial theoretical structure with input/output data to complete the model. The lateral \gls{hc} model is applied to capture the vehicle's response to both the left and right wheel tracks. 
To make the problem feasible, the parameters $m_s$, $m_u$, and $l$ are considered measurable and were not estimated. 
Thus, the system identification problem is reduced to estimating the parameters
$
\bm{\beta} =
\begin{bmatrix} 
\beta_1 & \beta_2 & \beta_3 & \beta_4
\end{bmatrix}
=
\begin{bmatrix} 
K_s & C_s & K_t & I_s
\end{bmatrix}.
$
The state-space system representing the \gls{hc} model, previously introduced in \eqref{eq:HC_sys} and parameterized by $\bm{\beta}$, becomes,
\begin{align*}\label{eq:sys_identification_HC_sys}
\setlength{\arraycolsep}{2pt}
\dot{\bm{x}}_{HC}(t) &=
A_{HC}
\bm{x}_{HC}(t)
+
B_{HC} \bm{u}(t),
\\
\hat{\bm{y}}_{HC}(t,\bm{\beta}) &= C_{HC}  \bm{x}_{HC}(t).
\end{align*}

Let $\bm{y}_{HC}(t)$ denote the measured vibration response, and let $\hat{\bm{y}}_{HC}(t,\bm{\beta})$ denote the estimated vibration response for a given road section. Also, let $\bm{Y}_{HC}(f)$ and $\hat{\bm{Y}}_{HC}(f,\bm{\beta})$ denote their respective Fourier transforms. The unknown parameters $\bm{\beta}$ are estimated iteratively by minimizing the difference between the amplitude spectrum $\lvert \bm{Y}_{HC}(f) \rvert$ and $\lvert \hat{\bm{Y}}_{HC}(f,\bm{\beta}) \rvert$:
\begin{equation*}
    \hat{\bm{\beta}} = \argmin_{\bm{\beta}}{\int_{0}^{F} \left( \lvert \bm{Y}_{HC}(f) \rvert - \lvert \hat{\bm{Y}}_{HC}(f,\bm{\beta}) \rvert \right)^2 \, df},
\end{equation*}
where $F$ is chosen such that the optimization covers the frequency range that most significantly impacts the \gls{iri}. The optimization is carried out using the \texttt{fminunc} function in MATLAB, configured to use the Quasi-Newton method, a gradient-based optimization algorithm.

\section{IRI Estimation Method} \label{sec:method_KF_based_method}
The proposed method presented for estimating the \gls{iri} aims first to estimate the longitudinal road profile from vibration measurements using a \gls{kf}, and then calculate the \gls{iri} values from the estimated profile using the ``Golden car'' parameters.

\subsection{Road Profile Estimation}
The state-space systems relating the road profile to the measured vibrations of the vehicle's sprung mass are described by the \gls{qc} and \gls{hc} models. Both systems have the structure,
\begin{align*}
    \dot{\bm{x}}(t) &= \bm{A} \bm{x}(t) + \bm{B} \bm{u}(t), \\
    \bm{y}(t) &= \bm{C} \bm{x}(t),
\end{align*}
where $\bm{x}(t)$ represents $\bm{x}_{QC}(t)$ or $\bm{x}_{HC}(t)$, $\bm{A}$ represents $\bm{A}_{QC}$ or $\bm{A}_{HC}$ etc. The system is discretized to be used in the \gls{kf},
\begin{align*}
    \bm{x}[k+1] &= \bm{F} \bm{x}[k] + \bm{G} \bm{u}[k] = e^{\bm{A}T} \bm{x}[k] + \int_0^T e^{\bm{A} \tau} \bm{B} \, d\tau \cdot \bm{u}[k], \\
    \bm{y}[k] &= \bm{H} \bm{x}[k],
\end{align*}
where $T$ is the sampling time and $\bm{H} = \bm{C}$.

To estimate the unknown input $\bm{u}[k]$, \ie, the road profile, the state vector $\bm{x}[k]$ is first estimated as $\hat{\bm{x}}[k]$ using the \gls{kf}, and the estimated input $\hat{\bm{u}}[k]$ is then obtained from $\hat{\bm{x}}[k]$. The complete algorithm is summarized as follows \cite{Kalman:Filter}.
The time update equations are given by,
\begin{align*}
\hat{\bm{x}}[k+1|k] &= \bm{F} \hat{\bm{x}}[k|k], \\
\bm{P}[k+1|k] &= \bm{F} \bm{P}[k|k] \bm{F}^T + \bm{G} \bm{Q} \bm{G}^T,
\end{align*}
where $\bm{P}[k|k]$ is the state covariance matrix and $\bm{Q}$ is the process noise covariance matrix. 
The measurement update equations are given by,
\begin{align*}
\bm{S} &= \bm{H} \bm{P}[k+1|k] \bm{H}^T + \bm{R}, \\
\bm{K}[k+1] &= \bm{P}[k+1|k] \bm{H}^T \bm{S}^{-1}, \\
\bm{\epsilon}[k+1] &= \left(\bm{y}[k+1] - \bm{H} \hat{\bm{x}}[k+1|k]\right), \\
\hat{\bm{x}}[k+1|k+1] &= \hat{\bm{x}}[k+1|k] + \bm{K}[k+1] \bm{\epsilon}[k+1], \\
\bm{P}[k+1|k+1] &= \left(\bm{I} - \bm{K}[k+1] \bm{H}\right) \bm{P}[k+1|k],
\end{align*}
where $\bm{R}$ is the measurement noise covariance matrix.

Using the estimated state $\hat{\bm{x}}[k+1|k+1]$, the input $\hat{\bm{u}}[k]$ is calculated as,
\begin{align*}
\hat{\bm{u}}[k|k\!+\!1] = \bm{Q} \bm{G}^T \left(\bm{P}[k\!+\!1|k]\right)^{-1} \! \left(\hat{\bm{x}}[k\!+\!1|k\!+\!1] - \bm{F}\hat{\bm{x}}[k|k]\right).
\end{align*}

Since the input $\bm{u}[k]$ is unknown, it is not used in the filter's time update. Instead, the input is modeled as a zero-mean process, $\mathbb{E}\left( \bm{u}[k] \right) = 0$, with process noise covariance $\bm{Q} = \text{Cov}(\bm{u}[k])$. This assumption simplifies the model by treating the road profile as a random process with known statistical properties. To handle the uncertainty associated with this assumption, the process noise covariance $\bm{Q}$ is set to a high value, which places greater trust in the measurement updates. As a result, $\hat{\bm{u}}[k]$ is the estimated road profile at time $k$ that, in the least squares sense, best explains the observed state transition from $\hat{\bm{x}}[k|k]$ to $\hat{\bm{x}}[k+1|k+1]$ under the given model dynamics. This approach accomplishes an estimation of the input without the direct inversion of the model.

\subsection{IRI Calculation}
Since the \gls{iri} is defined at a fixed vehicle speed, $V = 80~\text{km/h}$, it can be expressed in the spatial domain and approximated as a discrete calculation,
\begin{align*}
    IRI &= \frac{1}{L} \int_{0}^{L/V} \left| \xi(t) \right| \, dt    =
    \frac{1}{LV} \int_{0}^{L} \left| \xi(s(t)/V) \right| \, ds
    \\&\approx
    \frac{S}{LV} \sum_{i=1}^{L/S} |\xi(iS/V)|,
\end{align*}
where the variable change $s(t) = V t$ is introduced to express the integral in terms of the distance traveled, where $S$ is the spatial distance between each sample. 

The state-space system relating the road profile $u(t)$ to the rattle space derivative $\xi(t)$ is defined using the ``Golden car'' parameters from Table~\ref{tab:golden_car} as,
\begin{align*}
\dot{\bm{x}}_{QC}(t) &= \bm{A}_{QC} \bm{x}_{QC} + \bm{B}_{QC} u(t), \\
\xi(t) &= \bm{C}_{\xi} \bm{x}_{QC}(t),
\end{align*}
where $\bm{C}_{\xi} = \begin{bmatrix} 0 & 1 & 0 & -1 \end{bmatrix}$. The system can be discretized as,
\begin{align*}
    \bm{x}_{QC}[i+1] 
    &= e^{\bm{A}_{QC}S/V} \bm{x}_{QC}[i] + \int_0^{S/V} e^{\bm{A}_{QC} \tau} \bm{B}_{QC} d\tau \cdot \bm{u}[t_i], \\
    \xi[i] &= \bm{C}_{\xi} \bm{x}_{QC}[i],
\end{align*}
where $t_i$ is the solution to,
\begin{align*}
    iS &= \int_0^{t_i} v(t) \, dt \approx (i - 1)S + (t_i - t_{i-1})v(t_i) \\
    &\Rightarrow
    t_i \approx t_{i-1} + \frac{S}{v(t_i)},
\end{align*}
where $v(t)$ is the vehicle's speed and $kT \leq t_i \leq (k+1)T$.

To clarify, as the estimated road profile $\hat{\bm{u}}[k]$ is calculated at discrete time instances $k$ in the \gls{kf}, and when the vehicle has traveled at least the distance $S$ since the last \gls{iri} calculation, the road profile at time $t_i$ is interpolated as,
\begin{equation*}
\bm{u}[t_i] = \hat{\bm{u}}[k] + (\hat{\bm{u}}[k+1] - \hat{\bm{u}}[k]) \frac{(t_i-kT)}{T}.
\end{equation*}
This ensures that the \gls{iri} is calculated at equally spaced distances $S$, regardless of the vehicle's speed, and no further speed compensation is needed.

\subsection{Choice of Model}
The model used in the \gls{kf} can be either the \gls{qc} or \gls{hc} model. If vertical vibrations $\ddot{z}_s(t)$ are measured, using the \gls{qc} model is the natural choice.

The \gls{hc} model can capture how differences between the left and the right wheel tracks affect lateral vibrations $\ddot{x}_s(t)$. If the \gls{hc} model is employed, both tracks are estimated. However, if only lateral vibrations are used, the result becomes consistently zero if the estimated tracks are averaged sample by sample, since lateral vibrations only capture differences between the tracks. Therefore, if only lateral vibrations are used, the estimated road profile cannot be the average of the tracks, but rather must be one of the wheel tracks (the choice of left or right does not matter).

Another challenge when using lateral vibrations to estimate the road profile is the influence of centripetal acceleration on the lateral vibration measurements due to turning. The centripetal acceleration is a low frequency component that can be filtered out from the lateral vibrations. To mitigate its effect, the lateral vibration measurements $\ddot{x}_s(t)$ is high-pass filtered before being used to estimate \gls{iri}.

\section{Results}\label{sec:results}
\subsection{System Identification} \label{sec:results_system_identification}
The system was identified according to Section~\ref{sec:method_system_identification}. A $500~\text{meter}$ long road section with significantly high roughness was chosen for the system identification, in order to significantly excite the suspension system. The model parameters for the Audi A6 e-tron are identified.

When minimizing the difference between the amplitude spectra $\lvert \bm{Y}_{HC}(f) \rvert$ and $\lvert \hat{\bm{Y}}_{HC}(f, \bm{\beta}) \rvert$, it was observed that introducing a scaling factor $\mu$ to the estimated vibration response significantly improved the stability of the optimization process and reduced the cost function. 
The values of $K_s$, $C_s$, and $K_t$ are initialized with the ``Golden car'' parameter values listed in \Tabref{tab:golden_car}, while $I_s$ and $\mu$ are initialized to $3\,000~\text{kgm}^2$ and $1$, respectively. The estimated model parameters $
\bm{\beta} =
\begin{bmatrix}
K_s & C_s & K_t & I_s
\end{bmatrix}
$ 
and $\mu$ are obtained as, 
\begin{equation*}
    \bm{\beta}, \mu = \argmin_{\bm{\beta}, \mu}{\int_{0.5}^{15} \left( \lvert \bm{Y}_{HC}(f) \rvert - \mu \lvert \hat{\bm{Y}}_{HC}(f,\bm{\beta}) \rvert \right)^2 \, df}.
\end{equation*}

\begin{figure}[tb]
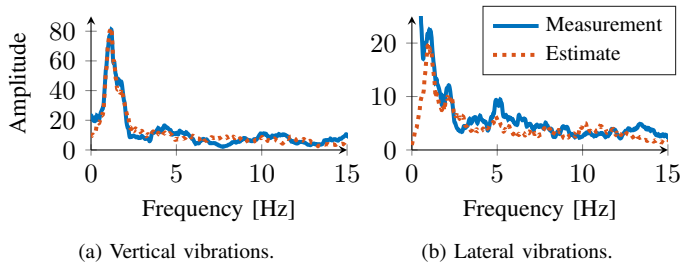

    \centering
    \text{Amplitude spectra obtained through system identification} 
    \\\hspace{-0.75cm}
    \subfloat[Vertical vibrations. \label{fig:sys_identification_freq_vert}]{%
        \setlength\figurewidth{0.4\columnwidth}
        \setlength\figureheight{0.2\columnwidth}
        \input{Figures/sys_identification_freq_vert.tex}
    } \hspace{-0.6cm}
    \subfloat[Lateral vibrations. \label{fig:sys_identification_freq_lat}]{%
        \setlength\figurewidth{0.4\columnwidth}
        \setlength\figureheight{0.2\columnwidth}
        \input{Figures/sys_identification_freq_lat.tex}
    }\hspace{-0.5cm}
    \caption{The optimization used for system identification aims to minimize the difference between the spectra $\lvert \bm{Y}_{HC}(f) \rvert$ (solid) and $\mu\lvert \hat{\bm{Y}}_{HC}(f, \bm{\beta}) \rvert$ (dashed). The spectra have been smoothed with a moving average filter over a frequency interval of $0.5~\text{Hz}$.}
    \label{fig:system_identification_freq}
\end{figure}

The amplitude spectrum $\lvert \bm{Y}_{HC}(f) \rvert$ and $\mu \lvert \hat{\bm{Y}}_{HC}(f,\bm{\beta}) \rvert$ after optimization are shown in \Figref{fig:system_identification_freq}. The optimization considers frequencies up to $15~\text{Hz}$, corresponding to the frequency range that mostly impacts the \gls{iri}, as seen in \Figref{fig:iri_freq_resp}. The frequency interval from $0$ to $0.5~\text{Hz}$ was excluded, as centripetal acceleration during turns significantly affects lateral vibration measurements in this frequency range.

\Tabref{tab:system_identification} lists the estimated model parameter values. The estimated parameter values are $127\%$ to $186\%$ higher than the ``Golden car'' parameters listed in \Tabref{tab:golden_car}, a result that is considered reasonable, given that the physical vehicle's sprung mass is approximated to be $140\%$ higher than that of the ``Golden car.''
\begin{table}[tb]
  \centering
  \caption{Identified model parameters for the physical car.}
  \label{tab:system_identification}
  \begin{tabular}{m{4.6cm} m{1cm} m{1cm} m{0.5cm}}
    \toprule
    {Property} & {Init. value}  & {Est. value} & {Unit}  \\
    \midrule
    Total Sprung Mass ($m_s$)$^\ast$ & - & 2\,400 & kg  \\
    Unsprung Mass ($m_u$)$^\ast$ & - & 90 & kg  \\
    Suspension Spring Stiffness ($K_s$) & 15\,825 & 37\,050 & N/m \\
    Suspension Damping Coefficient ($C_s$) & 1\,500 & 4\,290 & Ns/m  \\
    Tire Stiffness ($K_t$) & 163\,250 & 370\,600 & N/m  \\
    Unsprung Mass' Moment of Inertia ($I_s$) & 3\,000 & 1\,960 & $\text{kgm}^2$  \\
    Half-Axis Width ($l$)* & - & 1.0 & m \\
    Constant Gain ($\mu$) & 1 & 0.72 & - \\
    \bottomrule
    \multicolumn{4}{l}{\scriptsize $^\ast$Approximated. The other parameters are estimated through system identification.}
  \end{tabular}
  \vspace{-0.1cm}
\end{table}

The system identification was challenging with the available data. The inaccuracy of the \gls{gnss} matching algorithm, used to synchronize road profile measurements with sprung mass vibration data, is considered a significant source of error in this context. To make the parameter identification less sensitive to mismatched input/output data, only the amplitude spectra of measured and estimated sprung mass vibrations were considered. In this way, the phase information was neglected.

The parameters $m_s$, $m_u$, and $l$ were not estimated but rather approximated using educated guesses. When attempting to estimate all \gls{hc} model parameters simultaneously, the obtained parameters varied substantially depending on the selected road segment over which the system identification was performed. Additionally, some parameters could be estimated at unrealistic values. The reason could be that the suspension system is not sufficiently excited and the optimization process gets stuck in a local minima. 

Furthermore, real vehicle suspensions are nonlinear, which the linear models used in this paper do not account for. During system identification, it was found that the estimated vibrations aligned more closely with the measured data when scaled by $\mu = 0.72$, which suggests that the linear \gls{hc} model overestimates the actual response. This may be explained by advancements in vehicle's suspension that reduce vibrations beyond what the linear \gls{hc} model accounts for.

\subsection{IRI Estimation} \label{sec:results_field_test}
First, the \gls{iri} estimation algorithm was validated using simulated data, confirming that the developed method provides highly accurate estimates with negligible errors in the absence of measurement noise or model inaccuracies. For further details, the reader is referred to \cite{Agebjar_2024}.

The method was then validated on real-world data. A reference \gls{iri}, averaged over $40~\text{m}$, was calculated by the Swedish National Road and Transport Research Institute (VTI), and \gls{imu} data from the Audi A6 e-tron  was recorded at $50~\text{Hz}$. The \gls{imu} data and \gls{iri} reference were then matched using \gls{gnss}.

The validation began with an analysis of vertical vibrations by employing the \gls{qc} model within the \gls{kf}. The gravitational acceleration $9.82~\text{m/s}^2$ was subtracted from the vertical vibration measurements, and tuning parameters were adjusted to minimize the \gls{rmse} between the estimated \gls{iri} and the reference. The best results were achieved when the measured vertical vibration signal is low-pass filtered with a cutoff frequency of $13~\text{Hz}$. Additionally, the quotient $\frac{\bm{Q}}{\bm{R}}$ of the process noise covariance $\bm{Q}$ to the measurement noise covariance $\bm{R}$ was set to a large value of $10^9$.

\begin{figure}[tb]
    \centering
    \setlength\figurewidth{0.8\columnwidth}
    \setlength\figureheight{0.4\columnwidth}
    \text{\gls{iri} estimation on real-world data}\\
    \hspace{0cm}
    \input{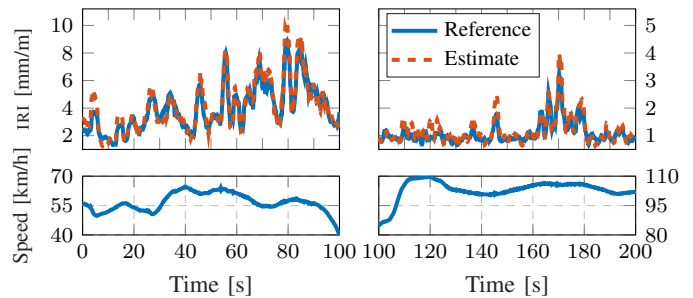}
    \caption{Estimated \gls{iri} on real-world validation data, using vertical vibrations. The first 100\,s have a higher roughness level and the vehicle travels at a lower speed, while the last 100\,s have a lower roughness level and the vehicle travels at a higher speed. The \gls{rmse} for the first segment is $0.48~\text{mm/m}$, while the \gls{rmse} for the second segment is $0.22~\text{mm/m}$.}
    \label{fig:iri_est_kf_field}
    \vspace{-0.1cm}
\end{figure}

The method was first validated on two road segments, covering a total distance of around $4.4~\text{km}$. One segment had high roughness where the vehicle traveled at around $55~\text{km/h}$, while the other had low roughness with a speed of approximately $105~\text{km/h}$. The estimated \gls{iri} is shown in \Figref{fig:iri_est_kf_field}, where the \gls{rmse} for the first segment is $0.48~\text{mm/m}$, while the \gls{rmse} for the second segment is $0.22~\text{mm/m}$.

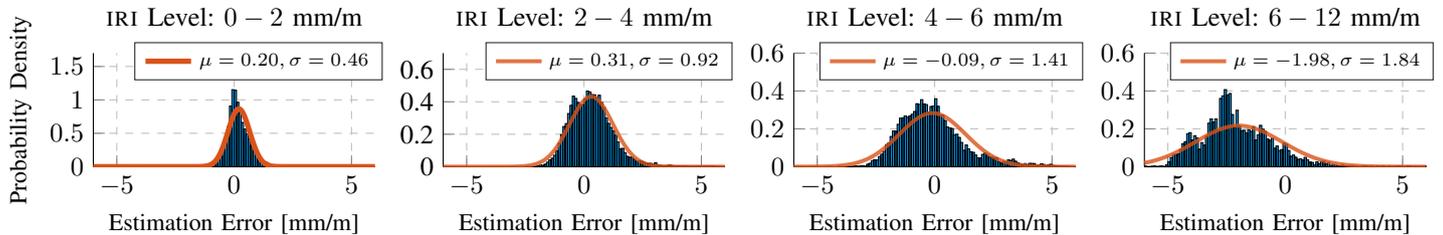
\begin{figure*}[tb!]
    \centering
    \text{\textbf{Vertical acceleration}}\\
    \hspace{-1.5cm} \hfill
    \subfloat{
        \setlength\figurewidth{0.44\columnwidth}
        \setlength\figureheight{0.17\columnwidth}
%
%
\definecolor{mycolor1}{rgb}{0.00000,0.54700,0.84100}%
\definecolor{mycolor2}{rgb}{0.85000,0.32500,0.09800}%
\begin{tikzpicture}

\begin{axis}[%
width=0.951\figurewidth,
height=\figureheight,
at={(0\figurewidth,0\figureheight)},
scale only axis,
bar shift auto,
xmin=-6,
xmax=6,
xlabel style={font=\small},
ylabel style={font=\small},
xlabel={},
ymin=0,
ymax=3,
ytick={ 0, 1, 2, 3},
yticklabels={{\color{white} 1.}0, 1, 2,3},
ylabel={Probability Density},
axis background/.style={fill=white},
axis x line*=bottom,
axis y line*=left,
xmajorgrids,
ymajorgrids,
grid style={dashed},
title = {\gls{iri} Level: $0-2~\text{mm/m}$},
legend style={legend cell align=left,font=\small, xshift=5,yshift=5, align=left, draw=white!15!black}
]
\addplot[forget plot, ybar interval, bar width=0.1, line width=0.2, fill=mycolor1, area legend] table[row sep=crcr] {%
x	y\\
-1.3	9.50139076607338e-05\\
-1.2	0.000451316061388486\\
-1.1	0.00339674719887123\\
-1	0.00408559802941156\\
-0.9	0.00966766510447967\\
-0.8	0.0225182961155939\\
-0.7	0.0431363140779732\\
-0.6	0.0983156409519444\\
-0.5	0.217866890266063\\
-0.4	0.423524493397721\\
-0.3	0.735977728740044\\
-0.2	1.26791309077866\\
-0.0999999999999999	1.99080265373844\\
0	2.02063702074391\\
0.1	1.4456841120119\\
0.2	0.783840984724138\\
0.3	0.39950972823647\\
0.4	0.220265991434497\\
0.5	0.114919321315657\\
0.6	0.070167770807452\\
0.7	0.0453928943849156\\
0.8	0.0276728006061887\\
0.9	0.0209505666391918\\
1	0.0130644123033509\\
1.1	0.00731607088987653\\
1.2	0.00351551458344715\\
1.3	0.00261288246067018\\
1.4	0.00194778510704504\\
1.5	0.00163898990714766\\
1.6	0.000926385599692158\\
1.7	0.000356302153727752\\
1.8	0.000380055630642935\\
1.9	0.000213781292236651\\
2	0.00021378129223665\\
2.1	0.000166274338406285\\
2.2	0.000118767384575918\\
2.3	7.126043074555e-05\\
2.4	9.50139076607342e-05\\
2.5	2.37534769151833e-05\\
2.6	2.37534769151835e-05\\
2.7	9.50139076607333e-05\\
2.8	7.12604307455506e-05\\
2.9	7.12604307455506e-05\\
3	4.75069538303667e-05\\
3.1	2.37534769151835e-05\\
3.2	7.126043074555e-05\\
3.3	0\\
3.4	4.75069538303667e-05\\
3.5	9.50139076607342e-05\\
3.6	9.50139076607342e-05\\
};
\addplot [color=mycolor2, line width=1.5pt, opacity=0.8]
  table[row sep=crcr]{%
  -6	0\\
-0.949999999999999	0.000753612094122502\\
-0.899999999999999	0.00163551463671929\\
-0.85	0.00340607848033603\\
-0.8	0.00680688644262162\\
-0.75	0.0130537713153718\\
-0.699999999999999	0.0240224409539289\\
-0.649999999999999	0.042422075122321\\
-0.6	0.0718886490804485\\
-0.55	0.116902117015806\\
-0.5	0.182422359708399\\
-0.449999999999999	0.273166477289117\\
-0.399999999999999	0.392527758865375\\
-0.35	0.541261313338434\\
-0.3	0.716204819231141\\
-0.25	0.909412942518101\\
-0.199999999999999	1.10809925287974\\
-0.149999999999999	1.29565636577617\\
-0.0999999999999996	1.4537664564516\\
-0.0499999999999998	1.56528381402552\\
0	1.61727988333245\\
0.0499999999999998	1.60350720046877\\
0.0999999999999996	1.52563373745432\\
0.149999999999999	1.3929107601017\\
0.199999999999999	1.22036555817182\\
0.25	1.02600678056905\\
0.3	0.82775947239514\\
0.35	0.640843121720971\\
0.399999999999999	0.476094248610111\\
0.449999999999999	0.33941246084559\\
0.5	0.23219682107463\\
0.55	0.152432789247078\\
0.6	0.0960271773011394\\
0.649999999999999	0.058050179809193\\
0.699999999999999	0.0336749233007305\\
0.75	0.0187457707203595\\
0.8	0.0100136780942623\\
0.85	0.00513307477998026\\
0.899999999999999	0.0025249640317373\\
1	0.000539872647341788\\
1.2	1.50482936698282e-05\\
5.3	0\\
6	0\\
};
\addlegendentry{\scriptsize $\mu = 0.01, \sigma = 0.25$}
\end{axis}
\end{tikzpicture}%
    } \hspace{-0.5cm}
    \subfloat{
        \setlength\figurewidth{0.44\columnwidth}
        \setlength\figureheight{0.17\columnwidth}
%
%
\definecolor{mycolor1}{rgb}{0.00000,0.54700,0.84100}%
\definecolor{mycolor2}{rgb}{0.85000,0.32500,0.09800}%
\begin{tikzpicture}

\begin{axis}[%
width=0.951\figurewidth,
height=\figureheight,
at={(0\figurewidth,0\figureheight)},
scale only axis,
bar shift auto,
xmin=-6,
xmax=6,
xlabel={},
ymin=0,
ymax=1.5,
ylabel={},
axis background/.style={fill=white},
axis x line*=bottom,
axis y line*=left,
xmajorgrids,
ymajorgrids,
grid style={dashed},
title = {\gls{iri} Level: $2-4~\text{mm/m}$},
legend style={legend cell align=left,font=\small, xshift=5,yshift=5, align=left, draw=white!15!black}
]
\addplot[ybar interval, bar width=0.1, line width=0.2, fill=mycolor1, area legend, forget plot] table[row sep=crcr] {%
x	y\\
-2.6	0.000379575065713933\\
-2.5	0\\
-2.4	0\\
-2.3	0\\
-2.2	0\\
-2.1	0.000379575065713933\\
-2	0.000379575065713933\\
-1.9	0.000474468832142417\\
-1.8	0.00142340649642725\\
-1.7	0.00474468832142416\\
-1.6	0.00635788235070838\\
-1.5	0.00474468832142416\\
-1.4	0.00873022651142048\\
-1.3	0.0118617208035604\\
-1.2	0.0161319402928422\\
-1.1	0.044979645287101\\
-1	0.0836014082234937\\
-0.9	0.133420635598448\\
-0.8	0.200700315996242\\
-0.7	0.257636575853333\\
-0.6	0.343420540704681\\
-0.5	0.529791897970222\\
-0.4	0.673081485277232\\
-0.3	0.833262163008511\\
-0.2	0.934988280619849\\
-0.1	0.945236807394121\\
0	0.968865355234814\\
0.1	0.908228238487013\\
0.2	0.718061130564333\\
0.3	0.548011501124493\\
0.4	0.408327876941763\\
0.5	0.376158890122507\\
0.6	0.292937056964728\\
0.7	0.230496958654786\\
0.8	0.151830026285574\\
0.9	0.117668270371319\\
1	0.0790465074349265\\
1.1	0.0521915715356658\\
1.2	0.0321689868192558\\
1.3	0.0187889657528398\\
1.4	0.0103434205407046\\
1.5	0.0119566145699889\\
1.6	0.00683235118285082\\
1.7	0.0113872519714179\\
1.8	0.00427021948928176\\
1.9	0.00417532572285324\\
2	0.00559873221928054\\
2.1	0.00493447585428111\\
2.2	0.00151830026285574\\
2.3	0.000284681299285451\\
2.4	0.000189787532856966\\
2.5	0.000189787532856966\\
};
\addplot [color=mycolor2, line width=1.5pt, opacity=0.8]
  table[row sep=crcr]{%
  -6	0\\
-1.8	0.000380445659468265\\
-1.65	0.00129222680215779\\
-1.55	0.00275405231982795\\
-1.5	0.00395066056943527\\
-1.45	0.0056012865321291\\
-1.4	0.00784921818441831\\
-1.35	0.0108714019545006\\
-1.3	0.0148821352065127\\
-1.25	0.0201356418796008\\
-1.2	0.0269268934243323\\
-1.15	0.0355899653025693\\
-1.1	0.0464931978076626\\
-1.05	0.0600304794153068\\
-1	0.0766081095944831\\
-0.949999999999999	0.0966269385454526\\
-0.899999999999999	0.120459826794101\\
-0.85	0.14842490824072\\
-0.8	0.180755651203122\\
-0.75	0.217569252356022\\
-0.699999999999999	0.258835413047588\\
-0.649999999999999	0.304347970899874\\
-0.6	0.353702123094327\\
-0.55	0.406280017946145\\
-0.5	0.4612472602436\\
-0.399999999999999	0.574000265656325\\
-0.35	0.629190346483776\\
-0.3	0.681667436856171\\
-0.25	0.729933987900596\\
-0.199999999999999	0.772529663271227\\
-0.149999999999999	0.808104048707808\\
-0.0999999999999996	0.835487470403087\\
-0.0499999999999998	0.853754764149999\\
0	0.862277153346323\\
0.0499999999999998	0.860758180218808\\
0.0999999999999996	0.849250826852805\\
0.149999999999999	0.8281544470864\\
0.199999999999999	0.798191757043873\\
0.25	0.760367733171468\\
0.3	0.715913677455483\\
0.35	0.666220789246184\\
0.399999999999999	0.612768231915671\\
0.449999999999999	0.557050852653657\\
0.55	0.444481476858459\\
0.6	0.390134079776911\\
0.649999999999999	0.338450103978658\\
0.699999999999999	0.290199028783773\\
0.75	0.245933560702348\\
0.8	0.205996661018066\\
0.85	0.170538761443208\\
0.899999999999999	0.139542522160787\\
0.949999999999999	0.112852340992636\\
1	0.0902059359100615\\
1.05	0.0712656419236648\\
1.1	0.055647521539484\\
1.15	0.0429469168491989\\
1.2	0.0327596080285266\\
1.25	0.0246982363552686\\
1.3	0.0184040637218246\\
1.35	0.0135544548242708\\
1.4	0.00986667722011791\\
1.45	0.00709872459671246\\
1.5	0.00504789455597976\\
1.55	0.00354781333440179\\
1.65	0.00169208960254519\\
1.75	0.000770137220214906\\
1.9	0.000216615385194352\\
2.25	7.4563160765706e-06\\
6	0\\
};
\addlegendentry{\scriptsize $\mu = 0.02, \sigma = 0.46$}
\end{axis}
\end{tikzpicture}%
    } \hspace{-0.5cm}
    \subfloat{
        \setlength\figurewidth{0.44\columnwidth}
        \setlength\figureheight{0.17\columnwidth}
%
%
\definecolor{mycolor1}{rgb}{0.00000,0.54700,0.84100}%
\definecolor{mycolor2}{rgb}{0.85000,0.32500,0.09800}%
\begin{tikzpicture}

\begin{axis}[%
width=0.951\figurewidth,
height=\figureheight,
at={(0\figurewidth,0\figureheight)},
scale only axis,
bar shift auto,
xmin=-6,
xmax=6,
xlabel={},
ymin=0,
ymax=0.9,
ytick={  0, .3, .6, .9},
ylabel={},
axis background/.style={fill=white},
axis x line*=bottom,
axis y line*=left,
xmajorgrids,
ymajorgrids,
grid style={dashed},
title = {\gls{iri} Level: $4-6~\text{mm/m}$},
legend style={legend cell align=left,font=\small, xshift=5,yshift=5, align=left, draw=white!15!black}
]
\addplot[ybar interval, bar width=0.1, line width=0.2, fill=mycolor1, area legend, forget plot] table[row sep=crcr] {%
x	y\\
-4.2	0.00145022115872671\\
-4.1	0\\
-4	0\\
-3.9	0\\
-3.8	0\\
-3.7	0\\
-3.6	0.00181277644840838\\
-3.5	0.00217533173809007\\
-3.4	0.00290044231745341\\
-3.3	0.00181277644840838\\
-3.2	0.00326299760713508\\
-3.1	0.00108766586904503\\
-3	0.00253788702777174\\
-2.9	0.000725110579363349\\
-2.8	0.0032629976071351\\
-2.7	0.00471321876586179\\
-2.6	0.00652599521427017\\
-2.5	0.00833877166267859\\
-2.4	0.00580088463490679\\
-2.3	0.00688855050395188\\
-2.2	0.00870132695236023\\
-2.1	0.0177652091944021\\
-2	0.0152273221666304\\
-1.9	0.013414545718222\\
-1.8	0.0130519904285404\\
-1.7	0.0286418678848524\\
-1.6	0.049307519396708\\
-1.5	0.0609092886665216\\
-1.4	0.100427815241824\\
-1.3	0.100065259952143\\
-1.2	0.122906243202088\\
-1.1	0.195054745848742\\
-1	0.276992241316801\\
-0.9	0.364730621419766\\
-0.8	0.407512145602206\\
-0.7	0.483648756435356\\
-0.6	0.520266840693205\\
-0.5	0.544920600391559\\
-0.4	0.593865564498586\\
-0.3	0.643535639184978\\
-0.2	0.595678340946991\\
-0.0999999999999996	0.596766006816042\\
0	0.558697701399465\\
0.0999999999999996	0.551084040316145\\
0.2	0.528968167645568\\
0.3	0.448843448625913\\
0.4	0.422014357189473\\
0.5	0.368718729606263\\
0.600000000000001	0.336088753534915\\
0.7	0.274091798999348\\
0.8	0.194692190559059\\
0.9	0.141034007686173\\
1	0.111304473932274\\
1.1	0.0699731709085638\\
1.2	0.0518454064244799\\
1.3	0.0282793125951706\\
1.4	0.0271916467261258\\
1.5	0.0163149880356754\\
1.6	0.0105141034007687\\
1.7	0.0108766586904503\\
1.8	0.0130519904285403\\
1.9	0.0126894351388587\\
2	0.00688855050395182\\
2.1	0.00543832934522517\\
2.2	0.00362555289681678\\
2.3	0.00108766586904502\\
2.4	0.00108766586904502\\
};
\addplot [color=mycolor2, line width=1.5pt, opacity=0.8]
  table[row sep=crcr]{%
  -6	0\\
-2.75	0.000255093193665346\\
-2.55	0.000797879689928394\\
-2.4	0.0017683356834155\\
-2.3	0.00292210914165381\\
-2.2	0.0047206192465028\\
-2.1	0.00745542370233032\\
-2.05	0.00929015725354265\\
-2	0.0115110914702194\\
-1.95	0.0141824953216547\\
-1.9	0.0173752674167069\\
-1.85	0.0211666949160358\\
-1.8	0.0256399578778517\\
-1.75	0.0308833404453654\\
-1.7	0.0369891145213099\\
-1.65	0.0440520684375949\\
-1.6	0.0521676627955276\\
-1.55	0.061429808167607\\
-1.5	0.0719282745928371\\
-1.45	0.0837457604495793\\
-1.4	0.0969546678316133\\
-1.35	0.111613652251976\\
-1.3	0.127764035423029\\
-1.25	0.145426189905898\\
-1.2	0.164596022356133\\
-1.15	0.185241696617663\\
-1.1	0.207300747745983\\
-1.05	0.230677741982452\\
-1	0.255242634751164\\
-0.949999999999999	0.28082996818393\\
-0.899999999999999	0.307239031135759\\
-0.8	0.361551670423685\\
-0.699999999999999	0.415944298265761\\
-0.649999999999999	0.442365891555401\\
-0.6	0.467811394750109\\
-0.55	0.491929268629837\\
-0.5	0.51437189869926\\
-0.449999999999999	0.534803831118055\\
-0.399999999999999	0.55291005731631\\
-0.35	0.568404063662046\\
-0.3	0.581035359921758\\
-0.25	0.590596210631917\\
-0.199999999999999	0.596927316822478\\
-0.149999999999999	0.599922231006201\\
-0.0999999999999996	0.599530334461297\\
-0.0499999999999998	0.595758260444711\\
0	0.588669707419413\\
0.0499999999999998	0.57838364964441\\
0.0999999999999996	0.565071015370479\\
0.149999999999999	0.548949962240404\\
0.199999999999999	0.53027993235958\\
0.25	0.509354713327724\\
0.3	0.486494764297764\\
0.35	0.46203908650462\\
0.399999999999999	0.436336925058437\\
0.449999999999999	0.409739583236054\\
0.649999999999999	0.30107783445522\\
0.699999999999999	0.274838937028952\\
0.75	0.24947121438101\\
0.8	0.225167303234194\\
0.85	0.202084459042052\\
0.899999999999999	0.180344627949632\\
0.949999999999999	0.16003545490917\\
1	0.141212099992872\\
1.05	0.12389971825491\\
1.1	0.108096449734429\\
1.15	0.0937767649086236\\
1.2	0.0808950163033462\\
1.25	0.0693890579931473\\
1.3	0.0591838101512234\\
1.35	0.0501946643450308\\
1.4	0.0423306456266976\\
1.45	0.0354972684317048\\
1.5	0.0295990438067131\\
1.55	0.0245416146500643\\
1.6	0.020233512782208\\
1.65	0.0165875462953737\\
1.7	0.0135218374938733\\
1.75	0.0109605407466287\\
1.8	0.00883427580961094\\
1.85	0.00708031584349733\\
1.95	0.00447140594055107\\
2.05	0.00276061844293451\\
2.15	0.00166624741323318\\
2.3	0.00074887516074007\\
2.5	0.000238177393034\\
2.9	1.83625322316416e-05\\
5.9	0\\
6	0\\
};
\addlegendentry{\scriptsize $\mu = -0.13, \sigma = 0.66$}
\end{axis}
\end{tikzpicture}%
    } \hspace{-0.62cm}
    \setcounter{subfigure}{0}
    \subfloat{
        \setlength\figurewidth{0.44\columnwidth}
        \setlength\figureheight{0.17\columnwidth}
%
%
\definecolor{mycolor1}{rgb}{0.00000,0.54700,0.84100}%
\definecolor{mycolor2}{rgb}{0.85000,0.32500,0.09800}%
\begin{tikzpicture}

\begin{axis}[%
width=0.951\figurewidth,
height=\figureheight,
at={(0\figurewidth,0\figureheight)},
scale only axis,
bar shift auto,
xmin=-6,
xmax=6,
xlabel={},
ymin=0,
ymax=0.9,
ytick={  0, .3, .6, .9},
ylabel={},
axis background/.style={fill=white},
axis x line*=bottom,
axis y line*=left,
xmajorgrids,
ymajorgrids,
grid style={dashed},
title = {\gls{iri} Level: $6-12~\text{mm/m}$},
legend style={legend cell align=left,font=\small, xshift=5,yshift=5, align=left, draw=white!15!black}
]
\addplot[ybar interval, bar width=0.1, line width=0.2, fill=mycolor1, area legend, forget plot] table[row sep=crcr] {%
x	y\\
-7.1	0.00113765642775882\\
-7	0.00796359499431168\\
-6.9	0.00227531285551764\\
-6.8	0.00568828213879405\\
-6.7	0\\
-6.6	0\\
-6.5	0.0056882821387941\\
-6.4	0.00113765642775882\\
-6.3	0.00341296928327643\\
-6.2	0.0102389078498294\\
-6.1	0.00341296928327643\\
-6	0\\
-5.9	0.0056882821387941\\
-5.8	0.00796359499431168\\
-5.7	0.0136518771331059\\
-5.6	0.018202502844141\\
-5.5	0.0102389078498294\\
-5.4	0.00227531285551764\\
-5.3	0\\
-5.2	0.00113765642775882\\
-5.1	0.00455062571103524\\
-5	0.00796359499431175\\
-4.9	0.0056882821387941\\
-4.8	0.00682593856655286\\
-4.7	0.00227531285551764\\
-4.6	0.00568828213879405\\
-4.5	0.00796359499431175\\
-4.4	0.0193401592719\\
-4.3	0.0580204778156993\\
-4.2	0.0250284414106941\\
-4.1	0.018202502844141\\
-4	0.012514220705347\\
-3.9	0.0102389078498293\\
-3.8	0.0216154721274175\\
-3.7	0.0204778156996588\\
-3.6	0.0216154721274175\\
-3.5	0.0375426621160409\\
-3.4	0.0352673492605233\\
-3.3	0.0500568828213879\\
-3.2	0.0602957906712175\\
-3.1	0.0147895335608645\\
-3	0.0307167235494882\\
-2.9	0.0307167235494882\\
-2.8	0.0216154721274174\\
-2.7	0.0250284414106941\\
-2.6	0.0591581342434582\\
-2.5	0.0739476678043234\\
-2.4	0.0386803185437996\\
-2.3	0.0295790671217293\\
-2.2	0.0295790671217293\\
-2.1	0.0489192263936289\\
-2	0.0466439135381117\\
-1.9	0.0329920364050055\\
-1.8	0.044368600682594\\
-1.7	0.0511945392491469\\
-1.6	0.0967007963594989\\
-1.5	0.0580204778156999\\
-1.4	0.0875995449374284\\
-1.3	0.118316268486917\\
-1.2	0.150170648464164\\
-1.1	0.19112627986348\\
-1	0.291240045506258\\
-0.9	0.254835039817974\\
-0.8	0.288964732650741\\
-0.7	0.301478953356088\\
-0.600000000000001	0.376564277588166\\
-0.5	0.41524459613197\\
-0.4	0.441410693970419\\
-0.3	0.440273037542664\\
-0.2	0.480091012514222\\
-0.100000000000001	0.455062571103524\\
0	0.534698521046646\\
0.0999999999999996	0.554038680318541\\
0.2	0.486916951080775\\
0.3	0.492605233219569\\
0.399999999999999	0.453924914675765\\
0.5	0.393629124004552\\
0.6	0.356086461888508\\
0.7	0.303754266211605\\
0.8	0.210466439135382\\
0.899999999999999	0.1740614334471\\
0.999999999999999	0.14220705346985\\
1.1	0.113765642775882\\
1.2	0.0864618885096704\\
1.3	0.0455062571103528\\
1.4	0.0364050056882823\\
1.5	0.0261660978384524\\
1.6	0.0170648464163823\\
1.7	0.0113765642775882\\
1.8	0.00455062571103528\\
1.9	0.00455062571103528\\
};
\addplot [color=mycolor2, line width=1.5pt, opacity=0.8]
  table[row sep=crcr]{%
-6	1.05773418948019e-05\\
-5.05	0.000259082138526878\\
-4.7	0.000725244440999973\\
-4.45	0.00144035619503402\\
-4.25	0.00242135270814003\\
-4.1	0.00351384304712621\\
-3.95	0.00502465836900612\\
-3.8	0.00707995661003125\\
-3.7	0.00882587306191063\\
-3.6	0.0109305057898545\\
-3.5	0.0134486391605879\\
-3.4	0.0164388686246344\\
-3.3	0.0199627794677832\\
-3.2	0.0240838322916304\\
-3.1	0.0288659383977397\\
-3	0.0343717201761242\\
-2.9	0.0406604663497578\\
-2.8	0.0477858091461432\\
-2.7	0.055793169548954\\
-2.65	0.060138887734186\\
-2.6	0.0647170368702801\\
-2.55	0.0695297603675114\\
-2.5	0.0745781688861111\\
-2.45	0.0798622560462467\\
-2.4	0.085380817411159\\
-2.3	0.0971101000432881\\
-2.2	0.109729650431527\\
-2.1	0.123179685428317\\
-2	0.137375626670106\\
-1.9	0.15220740558386\\
-1.8	0.167539564043799\\
-1.65	0.191120090194855\\
-1.45	0.222639666285756\\
-1.35	0.237949005503968\\
-1.25	0.252650840047008\\
-1.2	0.259700320551015\\
-1.15	0.266509748845767\\
-1.1	0.273050256954005\\
-1.05	0.279293580938962\\
-1	0.285212262871706\\
-0.949999999999999	0.29077985091567\\
-0.899999999999999	0.295971095500746\\
-0.85	0.300762139559961\\
-0.8	0.305130700829086\\
-0.75	0.309056244263887\\
-0.699999999999999	0.312520142710588\\
-0.649999999999999	0.315505824071963\\
-0.6	0.317998903342922\\
-0.55	0.319987298044009\\
-0.5	0.321461325756881\\
-0.449999999999999	0.322413782660125\\
-0.399999999999999	0.322840002174109\\
-0.35	0.322737893046835\\
-0.3	0.322107956445877\\
-0.25	0.320953281860941\\
-0.199999999999999	0.319279521863901\\
-0.149999999999999	0.317094846014866\\
-0.0999999999999996	0.314409874440186\\
-0.0499999999999998	0.311237591838043\\
0	0.307593242885924\\
0.0499999999999998	0.303494210228716\\
0.0999999999999996	0.2989598764136\\
0.149999999999999	0.294011471305638\\
0.199999999999999	0.288671906663824\\
0.25	0.282965599679455\\
0.3	0.27691828737563\\
0.35	0.270556833837332\\
0.399999999999999	0.2639090322855\\
0.5	0.249868996290736\\
0.6	0.235031456120304\\
0.699999999999999	0.219631744477891\\
0.899999999999999	0.188061481664216\\
1.05	0.164545723666156\\
1.15	0.149298510126235\\
1.25	0.134579790708605\\
1.35	0.120520166833135\\
1.45	0.107224764406086\\
1.55	0.0947732945601718\\
1.65	0.0832208900616997\\
1.7	0.0777927023684191\\
1.75	0.0725996012715937\\
1.8	0.0676423189513775\\
1.85	0.0629204203793128\\
1.9	0.0584323857238447\\
2	0.0501469183548275\\
2.1	0.0427553417364175\\
2.2	0.036215293997703\\
2.3	0.0304753813778209\\
2.4	0.0254777917895099\\
2.5	0.0211606946305167\\
2.6	0.0174603745875084\\
2.7	0.0143130666701357\\
2.8	0.0116564775798871\\
2.9	0.00943099407588388\\
3	0.00758059181624127\\
3.15	0.00539620123278262\\
3.3	0.00378506235410647\\
3.45	0.00261612093914998\\
3.65	0.00156248230610245\\
3.9	0.000790699409381368\\
4.25	0.000284458557034739\\
4.8	4.85193851060828e-05\\
6	5.14493902592505e-07\\
};
\addlegendentry{\scriptsize $\mu = -0.38, \sigma = 1.24$}
\end{axis}
\end{tikzpicture}%
\hspace{-0.5cm}
    }
    \\
    \text{\textbf{Lateral acceleration}}\\
    \hspace{-1.5cm} \hfill
    \subfloat{
        \setlength\figurewidth{0.44\columnwidth}
        \setlength\figureheight{0.17\columnwidth}
%
%
\definecolor{mycolor1}{rgb}{0.00000,0.54700,0.84100}%
\definecolor{mycolor2}{rgb}{0.85000,0.32500,0.09800}%
\begin{tikzpicture}

\begin{axis}[%
width=0.951\figurewidth,
height=\figureheight,
at={(0\figurewidth,0\figureheight)},
scale only axis,
bar shift auto,
xmin=-6,
xmax=6,
xlabel style={font=\small},
ylabel style={font=\small},
xlabel={Estimation Error [mm/m]},
ymin=0,
ymax=1.7,
ylabel={Probability Density},
axis background/.style={fill=white},
axis x line*=bottom,
axis y line*=left,
xmajorgrids,
ymajorgrids,
grid style={dashed},
title = {\gls{iri} Level: $0-2~\text{mm/m}$},
legend style={legend cell align=left,font=\small, xshift=5,yshift=5, align=left, draw=white!15!black}
]
\addplot[ybar interval, bar width=0.1, line width=0.2, fill=mycolor1, area legend, forget plot] table[row sep=crcr] {%
x	y\\
-1.3	9.50139076607338e-05\\
-1.2	0.00194778510704505\\
-1.1	0.00505949058293407\\
-1	0.0145608813490075\\
-0.9	0.0299293809131312\\
-0.8	0.0582910323498602\\
-0.7	0.0993845474131276\\
-0.6	0.157224263701599\\
-0.5	0.258746624037093\\
-0.4	0.361527918649092\\
-0.3	0.599894059492958\\
-0.2	0.907359064683092\\
-0.0999999999999999	1.15301752293992\\
0	1.14456128515811\\
0.1	0.965840124848276\\
0.2	0.807119392101018\\
0.3	0.662009401626163\\
0.4	0.564216337166353\\
0.5	0.487516360207225\\
0.6	0.41224159186301\\
0.7	0.318272837186543\\
0.8	0.25126427880881\\
0.9	0.186844849414833\\
1	0.135513585801122\\
1.1	0.0936362059996535\\
1.2	0.0801917380656593\\
1.3	0.0695976873614875\\
1.4	0.0483858324762287\\
1.5	0.0322809751277343\\
1.6	0.02135437574675\\
1.7	0.017743847255642\\
1.8	0.0160335969177488\\
1.9	0.0107603250425781\\
2	0.00679349439774243\\
2.1	0.00501198362910373\\
2.2	0.003753049352599\\
2.3	0.00266038941450053\\
2.4	0.00149646904565656\\
2.5	0.0009976460304377\\
2.6	0.000807618215116241\\
2.7	0.00118767384575917\\
2.8	0.000831371692031424\\
2.9	0.000973892553522525\\
3	0.000332548676812567\\
3.1	0.000190027815321468\\
3.2	0.000308795199897383\\
3.3	0.000498823015218854\\
3.4	0.000261288246067017\\
3.5	0.000142520861491101\\
3.6	9.50139076607342e-05\\
3.7	0.000403809107558117\\
3.8	0.000166274338406285\\
3.9	0.000332548676812567\\
4	0.000190027815321468\\
4.1	2.37534769151835e-05\\
4.2	4.75069538303667e-05\\
4.3	7.12604307455506e-05\\
4.4	7.12604307455506e-05\\
};
\addplot [color=mycolor2, line width=2.0pt]
  table[row sep=crcr]{%
-6	0\\
-1.6	0.000435473676989595\\
-1.45	0.00146698328314976\\
-1.35	0.00310848885853421\\
-1.3	0.0044458620606882\\
-1.25	0.00628433095784509\\
-1.2	0.00877927209285367\\
-1.15	0.0121214441703872\\
-1.1	0.0165404232226614\\
-1.05	0.022306696877906\\
-1	0.029731738858823\\
-0.949999999999999	0.0391653263169509\\
-0.899999999999999	0.0509893628167228\\
-0.85	0.0656075494229622\\
-0.8	0.0834304226125315\\
-0.75	0.104855560609357\\
-0.699999999999999	0.130243148235941\\
-0.649999999999999	0.159887569293601\\
-0.6	0.193986233504464\\
-0.55	0.23260739474836\\
-0.5	0.275659217729872\\
-0.449999999999999	0.322862732065136\\
-0.399999999999999	0.373731506376448\\
-0.35	0.427560819798967\\
-0.3	0.483428763270116\\
-0.199999999999999	0.59661043537163\\
-0.149999999999999	0.65120033976766\\
-0.0999999999999996	0.702481331649555\\
-0.0499999999999998	0.748947459906733\\
0	0.789158634719385\\
0.0499999999999998	0.82181425324121\\
0.0999999999999996	0.845822865640192\\
0.149999999999999	0.860362691736679\\
0.199999999999999	0.864928312665114\\
0.25	0.859359838429845\\
0.3	0.84385220394907\\
0.35	0.818943839860477\\
0.399999999999999	0.785485636310463\\
0.449999999999999	0.744592694583268\\
0.5	0.697582681210227\\
0.55	0.645905532790311\\
0.6	0.591069725402015\\
0.75	0.422110565075092\\
0.8	0.368537356474106\\
0.85	0.318004456454826\\
0.899999999999999	0.271194758713754\\
0.949999999999999	0.228573444688983\\
1	0.190399871378975\\
1.05	0.156748691397711\\
1.1	0.127537409905229\\
1.15	0.102557554792531\\
1.2	0.0815068517180553\\
1.25	0.0640201923916912\\
1.3	0.0496976928102715\\
1.35	0.0381286892251778\\
1.4	0.0289110536448609\\
1.45	0.0216656817484617\\
1.5	0.0160463845247083\\
1.55	0.0117456870846997\\
1.6	0.00849720417978617\\
1.65	0.00607533285867135\\
1.7	0.00429299612375988\\
1.8	0.00206933365945439\\
1.9	0.000951669456378923\\
2.05	0.0002717643673531\\
2.4	9.6727046274836e-06\\
6	0\\
};
\addlegendentry{\scriptsize $\mu = 0.20, \sigma = 0.46$}
\end{axis}
\end{tikzpicture}%
    }\hspace{-0.5cm}
    \subfloat{
        \setlength\figurewidth{0.44\columnwidth}
        \setlength\figureheight{0.17\columnwidth}
%
%
\definecolor{mycolor1}{rgb}{0.00000,0.54700,0.84100}%
\definecolor{mycolor2}{rgb}{0.85000,0.32500,0.09800}%
\begin{tikzpicture}

\begin{axis}[%
width=0.951\figurewidth,
height=\figureheight,
at={(0\figurewidth,0\figureheight)},
scale only axis,
bar shift auto,
xmin=-6,
xmax=6,
xlabel style={font=\small},
ylabel style={font=\small},
xlabel={Estimation Error [mm/m]},
ymin=0,
ymax=0.7,
ytick={  0, .2, .4, .6},
yticklabels={{\color{white} 1.}0, 0.2, 0.4, 0.6},
ylabel={},
axis background/.style={fill=white},
axis x line*=bottom,
axis y line*=left,
xmajorgrids,
ymajorgrids,
grid style={dashed},
title = {\gls{iri} Level: $2-4~\text{mm/m}$},
legend style={legend cell align=left,font=\small, xshift=5,yshift=5, align=left, draw=white!15!black}
]
\addplot[ybar interval, bar width=0.1, line width=0.2, fill=mycolor1, area legend, forget plot] table[row sep=crcr] {%
x	y\\
-2.5	0.000189787532856966\\
-2.4	0.000189787532856966\\
-2.3	0\\
-2.2	0.000189787532856966\\
-2.1	0.00161319402928421\\
-2	0.00446000702213871\\
-1.9	0.00920469534356287\\
-1.8	0.00768639508070716\\
-1.7	0.0156574714606998\\
-1.6	0.0275191922642601\\
-1.5	0.0414685759292472\\
-1.4	0.0522864653020943\\
-1.3	0.0705060684563632\\
-1.2	0.0979303669541947\\
-1.1	0.144238524971295\\
-1	0.197853503003388\\
-0.9	0.247577836611913\\
-0.8	0.265417864700468\\
-0.7	0.327478387944696\\
-0.6	0.377866977918221\\
-0.5	0.434708344008882\\
-0.4	0.420569172811038\\
-0.3	0.395707006006775\\
-0.2	0.384035072736072\\
-0.0999999999999996	0.380334215845362\\
0	0.417817253584612\\
0.1	0.466497755762424\\
0.2	0.460614342243858\\
0.3	0.443723251819588\\
0.4	0.443153889221019\\
0.5	0.423985348402463\\
0.6	0.440496863761019\\
0.7	0.39665594367106\\
0.8	0.363822700486805\\
0.9	0.300813239578293\\
1	0.266651483664038\\
1.1	0.241694423093347\\
1.2	0.221292263311223\\
1.3	0.193013920915535\\
1.4	0.152304495117716\\
1.5	0.110361450356325\\
1.6	0.113113369582753\\
1.7	0.0864482212163486\\
1.8	0.063673717273512\\
1.9	0.0579800912878035\\
2	0.0571260473899467\\
2.1	0.0457387954185291\\
2.2	0.0336872870821114\\
2.3	0.0368187813742517\\
2.4	0.0338770746149687\\
2.5	0.0276140860306885\\
2.6	0.0209715223806949\\
2.7	0.0199276909499814\\
2.8	0.020212372249267\\
2.9	0.0151830026285574\\
3	0.0246723792714055\\
3.1	0.0106281018399902\\
3.2	0.00835065144570649\\
3.3	0.00493447585428115\\
3.4	0.00446000702213873\\
3.5	0.00740171378142166\\
3.6	0.00967916417570533\\
3.7	0.00721192624856469\\
3.8	0.00379575065713935\\
3.9	0.00341617559142541\\
4	0.00446000702213869\\
4.1	0.00246723792714057\\
4.2	0.00313149429213993\\
4.3	0.00189787532856967\\
4.4	0.00123361896357029\\
4.5	0.00142340649642724\\
4.6	0.00161319402928422\\
4.7	0.00142340649642724\\
4.8	0.00199276909499816\\
4.9	0.00189787532856967\\
5	0.000948937664284828\\
5.1	0.00208766286142664\\
5.2	0.00246723792714055\\
5.3	0.000854043897856353\\
5.4	0.00265702545999754\\
5.5	0.00161319402928422\\
5.6	0.000189787532856964\\
5.7	0.00104383143071332\\
5.8	0\\
5.9	9.48937664284836e-05\\
6	0\\
6.1	9.4893766428482e-05\\
6.2	0.000284681299285451\\
6.3	0.000664256364999386\\
6.4	0.000379575065713935\\
6.5	0.000379575065713935\\
6.6	0.000189787532856964\\
6.7	0.000189787532856964\\
};
\addplot [color=mycolor2, line width=1.5pt, opacity=0.8]
  table[row sep=crcr]{%
-6	3.08268965909519e-11\\
-3.35	0.000166787116056888\\
-3.05	0.000573933289110151\\
-2.85	0.00123360009583529\\
-2.7	0.00212338537727863\\
-2.6	0.00300533945478954\\
-2.5	0.00420398331722804\\
-2.4	0.005812074537622\\
-2.3	0.00794152865110487\\
-2.25	0.00924227955311974\\
-2.2	0.0107245673068359\\
-2.15	0.0124081245124197\\
-2.1	0.0143139070146674\\
-2.05	0.0164640215363825\\
-2	0.0188816239753473\\
-1.95	0.0215907857845501\\
-1.9	0.0246163261656749\\
-1.85	0.0279836082195546\\
-1.8	0.031718297718176\\
-1.75	0.0358460837916637\\
-1.7	0.0403923615581103\\
-1.65	0.0453818775581034\\
-1.6	0.0508383397794132\\
-1.55	0.0567839950564455\\
-1.5	0.063239177685575\\
-1.45	0.0702218341889447\\
-1.4	0.0777470302597116\\
-1.35	0.0858264470013879\\
-1.3	0.0944678746005767\\
-1.25	0.103674712511594\\
-1.2	0.113445486047736\\
-1.15	0.12377338993173\\
-1.1	0.134645869822998\\
-1.05	0.146044253079633\\
-1	0.157943440000723\\
-0.949999999999999	0.170311666506387\\
-0.899999999999999	0.183110348632331\\
-0.85	0.196294018333332\\
-0.8	0.209810358905074\\
-0.75	0.223600346854473\\
-0.649999999999999	0.251733271920993\\
-0.5	0.294161262214123\\
-0.449999999999999	0.308024411955346\\
-0.399999999999999	0.321595881984281\\
-0.35	0.334781542428176\\
-0.3	0.347486725037246\\
-0.25	0.359617332354599\\
-0.199999999999999	0.371080982100366\\
-0.149999999999999	0.381788167642957\\
-0.0999999999999996	0.39165341439956\\
-0.0499999999999998	0.400596411420835\\
0	0.408543097301337\\
0.0499999999999998	0.415426679935268\\
0.0999999999999996	0.421188570509788\\
0.149999999999999	0.425779213484192\\
0.199999999999999	0.429158796117139\\
0.25	0.431297823335865\\
0.3	0.432177546338418\\
0.35	0.431790236218142\\
0.399999999999999	0.430139297025171\\
0.449999999999999	0.427239215951732\\
0.5	0.423115351660843\\
0.55	0.417803565083975\\
0.6	0.411349700205709\\
0.649999999999999	0.403808925349233\\
0.699999999999999	0.395244948198889\\
0.75	0.385729120176758\\
0.8	0.375339447772109\\
0.85	0.364159529960481\\
0.899999999999999	0.352277441911925\\
0.949999999999999	0.339784585758768\\
1	0.32677452926959\\
1.05	0.313341852870304\\
1.1	0.299581024590504\\
1.2	0.271445812401116\\
1.35	0.229022974543895\\
1.4	0.215143909460062\\
1.45	0.20151377846478\\
1.5	0.188194150994235\\
1.55	0.175239974767016\\
1.6	0.162699391043022\\
1.65	0.15061365852717\\
1.7	0.139017179739033\\
1.75	0.127937622080038\\
1.8	0.117396124528877\\
1.85	0.107407579895437\\
1.9	0.0979809818661499\\
1.95	0.0891198256734276\\
2	0.0808225511046068\\
2.05	0.0730830167110019\\
2.1	0.0658909944589468\\
2.15	0.0592326746515015\\
2.2	0.0530911717083393\\
2.25	0.0474470222869563\\
2.3	0.0422786682251459\\
2.35	0.0375629178477279\\
2.4	0.0332753802765167\\
2.45	0.0293908684807782\\
2.5	0.0258837678783852\\
2.55	0.0227283683217374\\
2.6	0.0198991582573873\\
2.65	0.0173710807185605\\
2.7	0.0151197515838382\\
2.75	0.0131216412056467\\
2.8	0.011354221075135\\
2.85	0.00979607764510693\\
2.95	0.00722801457277189\\
3.05	0.00527094592283461\\
3.15	0.00379892607120702\\
3.25	0.0027060502932299\\
3.35	0.00190508179141879\\
3.5	0.00110083671979222\\
3.7	0.000508504586148817\\
3.95	0.000181274930323383\\
4.4	2.35360229297044e-05\\
6	2.41774422704566e-09\\
};
\addlegendentry{\scriptsize $\mu = 0.31, \sigma = 0.92$}
\end{axis}
\end{tikzpicture}%
    }\hspace{-0.5cm}
    \subfloat{
        \setlength\figurewidth{0.44\columnwidth}
        \setlength\figureheight{0.17\columnwidth}
%
%
\definecolor{mycolor1}{rgb}{0.00000,0.54700,0.84100}%
\definecolor{mycolor2}{rgb}{0.85000,0.32500,0.09800}%
\begin{tikzpicture}

\begin{axis}[%
width=0.951\figurewidth,
height=\figureheight,
at={(0\figurewidth,0\figureheight)},
scale only axis,
bar shift auto,
xmin=-6,
xmax=6,
xlabel style={font=\small},
ylabel style={font=\small},
xlabel={Estimation Error [mm/m]},
ymin=0,
ymax=0.6,
ylabel={},
axis background/.style={fill=white},
axis x line*=bottom,
axis y line*=left,
xmajorgrids,
ymajorgrids,
grid style={dashed},
title = {\gls{iri} Level: $4-6~\text{mm/m}$},
legend style={legend cell align=left,font=\small, xshift=5,yshift=5, align=left, draw=white!15!black}
]
\addplot[ybar interval, bar width=0.1, line width=0.2, fill=mycolor1, area legend, forget plot] table[row sep=crcr] {%
x	y\\
-4.1	0.00145022115872671\\
-4	0\\
-3.9	0\\
-3.8	0\\
-3.7	0.000725110579363352\\
-3.6	0.000725110579363352\\
-3.5	0.000725110579363352\\
-3.4	0.00398810818649846\\
-3.3	0.00181277644840838\\
-3.2	0.00507577405554347\\
-3.1	0.00543832934522514\\
-3	0.00253788702777173\\
-2.9	0.0130519904285404\\
-2.8	0.0181277644840837\\
-2.7	0.0319048654919876\\
-2.6	0.036980639547531\\
-2.5	0.0351678630991226\\
-2.4	0.0532956275832066\\
-2.3	0.0656225074323831\\
-2.2	0.0804872743093324\\
-2.1	0.116380247987818\\
-2	0.125081574940178\\
-1.9	0.159524327459937\\
-1.8	0.189253861213835\\
-1.7	0.186715974186064\\
-1.6	0.223696613733594\\
-1.5	0.256326589804945\\
-1.4	0.258139366253353\\
-1.3	0.26829091436444\\
-1.2	0.261039808570808\\
-1.1	0.268653469654122\\
-1	0.276992241316801\\
-0.9	0.291494452904068\\
-0.8	0.332825755927779\\
-0.7	0.329200203030963\\
-0.600000000000001	0.354216518018998\\
-0.5	0.331738090058734\\
-0.4	0.328112537161917\\
-0.3	0.287143789427887\\
-0.2	0.317960989050831\\
-0.100000000000001	0.320136320788919\\
0	0.358929736784861\\
0.0999999999999996	0.319048654919876\\
0.199999999999999	0.270828801392211\\
0.3	0.271916467261258\\
0.399999999999999	0.243637154666085\\
0.5	0.20955695743601\\
0.6	0.204843738670146\\
0.7	0.201580741063013\\
0.8	0.16858820970198\\
0.899999999999999	0.171488652019432\\
1	0.154448553404395\\
1.1	0.127982017257631\\
1.2	0.0862881589442393\\
1.3	0.11456747153941\\
1.4	0.0939018200275537\\
1.5	0.0790370531506058\\
1.6	0.07577405554347\\
1.7	0.0616343992458852\\
1.8	0.0467696323689364\\
1.9	0.045681966499891\\
2	0.0648973968530203\\
2.1	0.0721485026466532\\
2.2	0.0638097309839753\\
2.3	0.0511202958451166\\
2.4	0.0507577405554344\\
2.5	0.044956855920528\\
2.6	0.046769632368936\\
2.7	0.0257414255673991\\
2.8	0.0184903197737656\\
2.9	0.0268290914364439\\
3	0.0239286491189907\\
3.1	0.0362555289681675\\
3.2	0.0322674207816693\\
3.3	0.0398810818649845\\
3.4	0.0286418678848523\\
3.5	0.0206656515118556\\
3.6	0.015589877456312\\
3.7	0.00870132695236027\\
3.8	0.0163149880356755\\
3.9	0.010151548111087\\
4	0.0174026539047202\\
4.1	0.0166775433253572\\
4.2	0.0163149880356755\\
4.3	0.00543832934522517\\
4.4	0.00942643753172362\\
4.5	0.0145022115872669\\
4.6	0.0116017692698137\\
4.7	0.00543832934522517\\
4.8	0.010151548111087\\
4.9	0.0126894351388587\\
5	0.00652599521427008\\
5.1	0.00471321876586181\\
5.2	0.00253788702777174\\
5.3	0.00108766586904503\\
5.4	0.00181277644840836\\
5.5	0.000725110579363356\\
5.6	0.00145022115872671\\
5.7	0.000362555289681678\\
5.8	0\\
5.9	0\\
6	0.00108766586904503\\
6.1	0.00253788702777174\\
6.2	0.00253788702777174\\
6.3	0.00145022115872671\\
6.4	0.00108766586904501\\
6.5	0.000725110579363356\\
6.6	0\\
6.7	0.000362555289681678\\
6.8	0\\
6.9	0.000725110579363343\\
7	0.000725110579363356\\
7.1	0.000725110579363356\\
7.2	0\\
7.3	0.000725110579363356\\
7.4	0.00145022115872669\\
7.5	0.000725110579363356\\
7.6	0.00145022115872671\\
7.7	0.000362555289681678\\
7.8	0.000362555289681678\\
};
\addplot [color=mycolor2, line width=1.5pt, opacity=0.8]
  table[row sep=crcr]{%
-6	4.26814298650768e-05\\
-5.35	0.000265935488426905\\
-5	0.00065207598230721\\
-4.75	0.00119154741464644\\
-4.55	0.00188673863424338\\
-4.35	0.00292790192013648\\
-4.2	0.00401739215947305\\
-4.05	0.00545013033209329\\
-3.95	0.00663711271610801\\
-3.85	0.00804197262677953\\
-3.75	0.00969520686316017\\
-3.65	0.0116295433350713\\
-3.55	0.0138796762089921\\
-3.45	0.0164818933793542\\
-3.35	0.0194735871269991\\
-3.25	0.0228926415212714\\
-3.15	0.0267766936185092\\
-3.05	0.0311622697877754\\
-2.95	0.0360838034885225\\
-2.85	0.0415725464175658\\
-2.75	0.0476553909644464\\
-2.7	0.0509265084054782\\
-2.65	0.0543536281433994\\
-2.6	0.0579383268422804\\
-2.55	0.0616816713376043\\
-2.5	0.0655841794798091\\
-2.45	0.0696457819893173\\
-2.4	0.0738657856855314\\
-2.35	0.0782428384629936\\
-2.25	0.0874591933518305\\
-2.15	0.0972696671530615\\
-2.05	0.107636727194205\\
-1.95	0.118509899691973\\
-1.85	0.129825459680314\\
-1.75	0.141506438157854\\
-1.65	0.153462978189196\\
-1.25	0.201847898881232\\
-1.15	0.213455385164893\\
-1.05	0.224595520859857\\
-1	0.229946729900479\\
-0.949999999999999	0.235128978879002\\
-0.899999999999999	0.240125261495257\\
-0.85	0.244918909369469\\
-0.8	0.249493684039703\\
-0.75	0.253833868120794\\
-0.699999999999999	0.25792435491295\\
-0.649999999999999	0.261750735748248\\
-0.6	0.26529938437055\\
-0.55	0.268557537658842\\
-0.5	0.271513372025868\\
-0.449999999999999	0.274156074852728\\
-0.399999999999999	0.276475910355919\\
-0.35	0.278464279325524\\
-0.3	0.280113772221759\\
-0.25	0.281418215171119\\
-0.199999999999999	0.282372708462603\\
-0.149999999999999	0.282973657208093\\
-0.0999999999999996	0.283218793898416\\
-0.0499999999999998	0.283107192657022\\
0	0.282639275065907\\
0.0499999999999998	0.281816807512462\\
0.0999999999999996	0.280642890080623\\
0.149999999999999	0.279121937084096\\
0.199999999999999	0.277259649412713\\
0.25	0.275062978934349\\
0.3	0.272540085263464\\
0.35	0.269700285272472\\
0.399999999999999	0.266553995783072\\
0.449999999999999	0.263112669930744\\
0.5	0.259388727746225\\
0.55	0.255395481542438\\
0.6	0.251147056733568\\
0.649999999999999	0.246658308744474\\
0.699999999999999	0.241944736693067\\
0.75	0.237022394545567\\
0.8	0.23190780045454\\
0.85	0.226617844992336\\
0.949999999999999	0.215580721665735\\
1.05	0.204050108267126\\
1.15	0.192165238709122\\
1.3	0.173971307562385\\
1.5	0.14969944159262\\
1.6	0.13781868278631\\
1.7	0.126242940371136\\
1.8	0.115058100473462\\
1.9	0.104337010852884\\
2	0.09413923681221\\
2.1	0.0845111570844779\\
2.15	0.0799217334306288\\
2.2	0.0754863653462197\\
2.25	0.0712073637577495\\
2.3	0.0670863360795249\\
2.35	0.063124218214103\\
2.4	0.0593213088243081\\
2.45	0.0556773054942212\\
2.5	0.0521913424028071\\
2.55	0.0488620291432911\\
2.65	0.042665405681416\\
2.75	0.0370673341057355\\
2.85	0.0320418746960236\\
2.95	0.0275584999726348\\
3.05	0.0235832863477778\\
3.15	0.0200800211337615\\
3.25	0.017011205811265\\
3.35	0.014338942539565\\
3.45	0.0120256965781831\\
3.55	0.010034932405862\\
3.65	0.00833162574132551\\
3.75	0.00688265731581339\\
3.85	0.0056570970916674\\
4	0.00417585739100712\\
4.15	0.00304770299671997\\
4.3	0.00219924990835629\\
4.5	0.00139856501021107\\
4.75	0.000772022836042829\\
5.05	0.000363001685726339\\
5.5	0.000107495073989661\\
6	2.46693364021056e-05\\
};
\addlegendentry{\scriptsize $\mu = -0.09, \sigma = 1.41$}
\end{axis}
\end{tikzpicture}%
    } \hspace{-0.5cm}
    \subfloat{
        \setlength\figurewidth{0.44\columnwidth}
        \setlength\figureheight{0.17\columnwidth}
%
%
\definecolor{mycolor1}{rgb}{0.00000,0.54700,0.84100}%
\definecolor{mycolor2}{rgb}{0.85000,0.32500,0.09800}%
\begin{tikzpicture}

\begin{axis}[%
width=0.951\figurewidth,
height=\figureheight,
at={(0\figurewidth,0\figureheight)},
scale only axis,
bar shift auto,
xmin=-6,
xmax=6,
xlabel style={font=\small},
ylabel style={font=\small},
xlabel={Estimation Error [mm/m]},
ymin=0,
ymax=0.6,
ylabel={},
axis background/.style={fill=white},
axis x line*=bottom,
axis y line*=left,
xmajorgrids,
ymajorgrids,
grid style={dashed},
title = {\gls{iri} Level: $6-12~\text{mm/m}$},
legend style={legend cell align=left,font=\small, xshift=5,yshift=5, align=left, draw=white!15!black}
]
\addplot[ybar interval, bar width=0.1, line width=0.2, fill=mycolor1, area legend, forget plot] table[row sep=crcr] {%
x	y\\
-6.1	0.00341296928327646\\
-6	0.00227531285551762\\
-5.9	0\\
-5.8	0\\
-5.7	0.00227531285551764\\
-5.6	0.00341296928327643\\
-5.5	0\\
-5.4	0.00227531285551764\\
-5.3	0.00341296928327643\\
-5.2	0\\
-5.1	0.0102389078498293\\
-5	0.0261660978384529\\
-4.9	0.0477815699658705\\
-4.8	0.0637087599544934\\
-4.7	0.0659840728100116\\
-4.6	0.0671217292377698\\
-4.5	0.0910125142207057\\
-4.4	0.121729237770194\\
-4.3	0.131968145620022\\
-4.2	0.1740614334471\\
-4.1	0.179749715585892\\
-4	0.141069397042094\\
-3.9	0.163822525597269\\
-3.8	0.151308304891923\\
-3.7	0.0921501706484645\\
-3.6	0.127417519908987\\
-3.5	0.113765642775882\\
-3.4	0.160409556313993\\
-3.3	0.229806598407281\\
-3.2	0.252559726962458\\
-3.1	0.235494880546075\\
-3	0.252559726962457\\
-2.9	0.189988623435722\\
-2.8	0.360637087599545\\
-2.7	0.374288964732652\\
-2.6	0.407281001137656\\
-2.5	0.378839590443686\\
-2.4	0.386803185437997\\
-2.3	0.275312855517633\\
-2.2	0.208191126279864\\
-2.1	0.298065984072808\\
-2	0.215017064846417\\
-1.9	0.189988623435723\\
-1.8	0.192263936291239\\
-1.7	0.178612059158135\\
-1.6	0.195676905574515\\
-1.5	0.186575654152447\\
-1.4	0.211604095563139\\
-1.3	0.141069397042094\\
-1.2	0.168373151308306\\
-1.1	0.218430034129692\\
-1	0.197952218430035\\
-0.9	0.186575654152445\\
-0.8	0.146757679180888\\
-0.7	0.138794084186576\\
-0.600000000000001	0.150170648464163\\
-0.5	0.0967007963594998\\
-0.4	0.109215017064846\\
-0.3	0.0853242320819116\\
-0.2	0.0898748577929468\\
-0.100000000000001	0.12400455062571\\
0	0.118316268486917\\
0.0999999999999996	0.0853242320819108\\
0.2	0.0546075085324234\\
0.3	0.0614334470989763\\
0.399999999999999	0.0500568828213877\\
0.5	0.0386803185437999\\
0.6	0.042093287827076\\
0.7	0.0352673492605234\\
0.8	0.0489192263936293\\
0.899999999999999	0.0364050056882819\\
1	0.0238907849829352\\
1.1	0.0227531285551762\\
1.2	0.0102389078498294\\
1.3	0.0250284414106941\\
1.4	0.0159271899886234\\
1.5	0.0273037542662117\\
1.6	0.0284414106939703\\
1.7	0.0147895335608647\\
1.8	0.0216154721274176\\
1.9	0.0056882821387941\\
2	0.011376564277588\\
2.1	0.00910125142207057\\
2.2	0.00341296928327646\\
2.3	0.00341296928327646\\
2.4	0.00910125142207057\\
2.5	0.0034129692832764\\
2.6	0.00455062571103528\\
2.7	0.00455062571103528\\
2.8	0.00455062571103528\\
2.9	0.00113765642775882\\
3	0.0011376564277588\\
3.1	0\\
3.2	0.00113765642775882\\
3.3	0.00227531285551764\\
3.4	0.0022753128555176\\
3.5	0\\
3.6	0.00113765642775882\\
3.7	0\\
3.8	0\\
3.9	0.0022753128555176\\
4	0\\
4.1	0.00227531285551764\\
4.2	0.00113765642775882\\
4.3	0\\
4.4	0.0022753128555176\\
4.5	0\\
4.6	0.00113765642775882\\
4.7	0.00113765642775882\\
4.8	0.00341296928327646\\
4.9	0.0011376564277588\\
5	0\\
5.1	0.00113765642775882\\
5.2	0.00341296928327646\\
5.3	0.00227531285551764\\
5.4	0\\
5.5	0.00113765642775882\\
5.6	0\\
5.7	0.00113765642775882\\
5.8	0\\
5.9	0.0011376564277588\\
6	0.00227531285551764\\
6.1	0.00113765642775882\\
6.2	0\\
6.3	0.00113765642775882\\
6.4	0.0011376564277588\\
6.5	0\\
6.6	0.00113765642775882\\
6.7	0.00341296928327646\\
6.8	0.00113765642775882\\
6.9	0.0011376564277588\\
7	0.00113765642775882\\
7.1	0.00113765642775882\\
7.2	0.00455062571103528\\
7.3	0.00113765642775882\\
7.4	0.0011376564277588\\
7.5	0.0113765642775882\\
7.6	0.00796359499431175\\
7.7	0.0295790671217293\\
7.8	0.0147895335608647\\
7.9	0.0136518771331057\\
8	0.022753128555176\\
8.1	0.00227531285551768\\
8.2	0.0022753128555176\\
8.3	0.0022753128555176\\
};
\addplot [color=mycolor2, line width=1.5pt, opacity=0.8]
  table[row sep=crcr]{%
-6	0.0198378844788767\\
-5.9	0.0223098901713001\\
-5.8	0.0250158766939883\\
-5.7	0.0279672807848241\\
-5.6	0.0311746063039493\\
-5.5	0.0346471831957444\\
-5.4	0.0383929167376742\\
-5.3	0.0424180306461812\\
-5.2	0.0467268081462713\\
-5.1	0.0513213356055902\\
-5	0.0562012537671057\\
-4.9	0.0613635219685822\\
-4.8	0.0668022009937035\\
-4.7	0.0725082603422678\\
-4.6	0.078469415720277\\
-4.5	0.0846700024226177\\
-4.35	0.0943770100526065\\
-4.2	0.104499533391024\\
-4.05	0.114940738395282\\
-3.8	0.132735717349852\\
-3.55	0.150479726700752\\
-3.4	0.160812857439171\\
-3.3	0.167472907584131\\
-3.2	0.17389398971877\\
-3.1	0.180028309344306\\
-3	0.185828899136403\\
-2.9	0.191250211917828\\
-2.8	0.196248712333309\\
-2.7	0.200783456556948\\
-2.6	0.204816649286172\\
-2.5	0.208314167454171\\
-2.4	0.211246040527601\\
-2.3	0.213586877942511\\
-2.2	0.215316235156092\\
-2.15	0.215946646868128\\
-2.1	0.216418910934637\\
-2.05	0.216731982980469\\
-2	0.216885169831964\\
-1.95	0.216878132074251\\
-1.9	0.216710885304334\\
-1.85	0.216383800073503\\
-1.8	0.215897600519347\\
-1.75	0.215253361694452\\
-1.65	0.213496795983707\\
-1.55	0.211129539644609\\
-1.45	0.208172261166929\\
-1.35	0.204650560382211\\
-1.25	0.200594599552656\\
-1.15	0.196038673427563\\
-1.05	0.191020726835085\\
-0.949999999999999	0.185581829291601\\
-0.85	0.179765616786359\\
-0.75	0.173617711322658\\
-0.649999999999999	0.167185128963529\\
-0.5	0.157107161860737\\
-0.35	0.146658019691067\\
-0.0999999999999996	0.1288457124327\\
0.149999999999999	0.111124674032363\\
0.3	0.100786758321139\\
0.449999999999999	0.0908046206458559\\
0.55	0.0843929570138595\\
0.649999999999999	0.0782025077396424\\
0.75	0.0722522499142224\\
0.85	0.0665576976970659\\
0.949999999999999	0.0611309895677401\\
1.05	0.0559810166350942\\
1.15	0.051113586594675\\
1.25	0.0465316176529296\\
1.35	0.0422353566150839\\
1.45	0.0382226153529279\\
1.55	0.0344890200166237\\
1.65	0.0310282676159979\\
1.75	0.027832384954702\\
1.85	0.0248919853369873\\
1.95	0.02219651896326\\
2.05	0.0197345134684772\\
2.15	0.017493801619211\\
2.3	0.0145199048260647\\
2.45	0.0119716727508088\\
2.6	0.00980522126677208\\
2.75	0.00797758526941106\\
2.9	0.00644758395657519\\
3.05	0.00517647417512723\\
3.25	0.00382287603663922\\
3.45	0.00279004532592264\\
3.7	0.00185104211249598\\
4	0.00110405615171683\\
4.35	0.000584196345355892\\
4.8	0.000244364324547774\\
5.5	5.59184538744262e-05\\
6	1.78465447389087e-05\\
};
\addlegendentry{\scriptsize $\mu = -1.98, \sigma = 1.84$}
\end{axis}
\end{tikzpicture}%
\hspace{-0.5cm}
    }
    \caption{Estimated \gls{iri} on real-world data, with vertical vibrations in the upper part and lateral vibrations in the lower part of the figure. The data covers over $230~\text{km}$ of driving, divided into different \gls{iri} levels. A standard distribution is fitted to each histogram and its mean value, $\mu$, and standard deviation, $\sigma$, are presented. Note that the $y$-axis is scaled differently for each subplot.}
    \label{fig:iri_est_kf_histogram_all}
\end{figure*}

The method was then validated on NIRA Dynamics' entire dataset for the specific car, consisting of calculated \gls{iri} references and \gls{imu} data, covering a total driven distance of over $230~\text{km}$. The results, categorized by different roughness levels, are presented as histograms in \Figref{fig:iri_est_kf_histogram_all}.

\begin{figure}[tb]
\centering
\setlength\figurewidth{0.6\columnwidth}
\setlength\figureheight{0.3\columnwidth}
\input{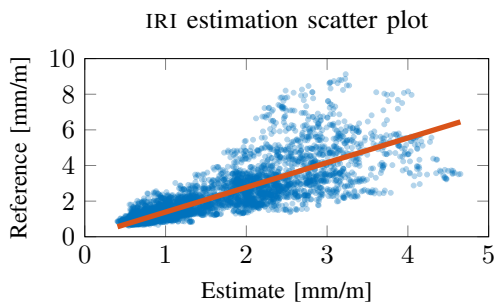}
\caption{Estimated \gls{iri} on real-world validation data using lateral vibrations. There is a large spread in the estimates. A regression line $Reference \approx 1.39 \cdot Estimate$ is fitted to the data.}
\label{fig:iri_est_kf_field_scatter}
\vspace{-0.1cm}
\end{figure}

The \gls{iri} was also estimated purely based on lateral vibrations by only considering measurements of $\ddot{x}_s(t)$ when employing the \gls{hc} model within the \gls{kf}. To reduce the impact of centripetal accelerations, the resulting lateral vibration signal was high-pass filtered with a cutoff frequency of $0.5~\text{Hz}$. The best result was obtained when the lateral vibration signal was low-pass filtered with a cutoff frequency of $11~\text{Hz}$. Other parameters were kept the same as those used for vertical vibrations. The results obtained when testing on the validation data are illustrated as a scatter plot in \Figref{fig:iri_est_kf_field_scatter}.

The scatter plot reveals a significant spread in the estimates, along with an overall attenuation. Linear regression analysis indicates that amplifying the estimate by 39\% minimizes the \gls{rmse} between the estimate and the reference. When validated on the entire dataset, the estimates is adjusted accordingly, and the results are presented as histograms in \Figref{fig:iri_est_kf_histogram_all}.
The results obtained throughout the field tests are summarized in \Tabref{tab:field_test_results}.
\begin{table}[tb]
\centering
\caption{Summary of \gls{iri} estimation results based on field tests.
The mean, standard deviation and \gls{rmse} of the estimation error are presented for different \gls{iri} levels. The best results are highlighted in bold.}
\label{tab:field_test_results}
\begin{tabular}{m{1.1cm} m{1cm} m{1cm} m{0.8cm} m{1cm} m{0.8cm}}
    \toprule
    \gls{iri} Level [mm/m] & Distance [km] & \centering Vibration Type & Mean [mm/m] & Std Dev [mm/m]& \gls{rmse} [mm/m]\\
    \midrule
    \multirow{2}{*}{0 -- 2}  & \multirow{2}{*}{190.9}  & \centering Vertical  & \textbf{0.01} & \textbf{0.25} & \textbf{0.25} \\
                             & & \centering Lateral   & 0.20 & 0.46 & 0.50\\
    \midrule
    \multirow{2}{*}{2 -- 4}  & \multirow{2}{*}{30.2}  & \centering Vertical  & \textbf{0.02} & \textbf{0.46} & \textbf{0.46}\\
                             & & \centering Lateral   & 0.31 & 0.92 & 0.97\\
    \midrule
    \multirow{2}{*}{4 -- 6}  & \multirow{2}{*}{7.3}  & \centering Vertical  & \textbf{-0.13} & \textbf{0.66} & \textbf{0.68}\\
                             & & \centering Lateral   & -0.09 & 1.41 & 1.41\\
    \midrule
    \multirow{2}{*}{6 -- 12} & \multirow{2}{*}{2.4}  & \centering Vertical  & \textbf{-0.38} & \textbf{1.24} & \textbf{1.29}\\
                             & & \centering Lateral   & -1.98 & 1.84 & 2.70\\ 
    \midrule[0.8pt]
    \multirow{2}{*}{0 -- 12} & \multirow{2}{*}{230.8}  & \centering Vertical  & \textbf{0.00} & \textbf{0.37} & \textbf{0.37}\\
                            & & \centering Lateral   & 0.17 & 0.74 & 0.76\\ 
    \bottomrule
\end{tabular}
\vspace{-0.1cm}
\end{table}

The method based on vertical acceleration provides the best results for all \gls{iri} levels. The mean estimation error is close to zero for all levels, and the standard deviation is below $1~\text{mm/m}$ for all levels except the highest. The method based on lateral acceleration provides worse results, with a mean estimation error close to zero only for the lowest \gls{iri} levels. The standard deviation is also higher for the lateral acceleration-based method, especially for the higher \gls{iri} levels.

It should be highlighted that the estimation errors illustrated in the histograms are the sum of several components: errors due to the vehicle not driving in the exact same wheel tracks that were laser-scanned, \gls{gnss} matching uncertainty, and the method's own estimation errors such as model errors and long sample intervals at high speeds.

The \glsxtrlong{kf} used in the method can account for measurement and process noise. However, since no input is provided in the filter's time update, much more trust is put in the measurement update. 
This limits the filter's ability to adapt to measurement noise, making the method quite sensitive to noise. 
Nonetheless, the method offers several advantages. 
The method can be implemented efficiently online in real-time. 
It also enables the estimation of \gls{iri} purely based on lateral vibrations by employing the \gls{hc} model. 
However, since lateral vibrations only capture differences between the left and right wheel tracks, it is impossible to obtain an estimate of the road's complete frequency content. 
Therefore, the \gls{iri} estimates based on lateral vibrations is significantly less accurate than the estimates obtained from vertical vibrations.
\section{Conclusions} \label{cha:conclusions}
This paper presented a model-based method for estimating road roughness in terms of the \gls{iri} by fusing measurements from a vehicle's onboard sensors. The approach leverages chassis vibrations and vehicle speed to estimate the road's longitudinal profile using a \glsxtrlong{kf}, which is then used to compute \gls{iri} values. The method can utilize both vertical and lateral sprung mass vibrations, and is straightforward to implement for real-time applications.

The method was validated on over 230~km of real-world data collected by NIRA Dynamics AB. The results show that, when using vertical vibrations, the method yields an estimation error between 1\% and 10\% relative to the reference values. However, when relying solely on lateral vibrations, accuracy significantly declines. To address this, a regression line was applied to the lateral vibration-based \gls{iri} estimates to correct the bias. Still, accurately estimating high road roughness levels with lateral vibrations remains challenging. This result is because lateral vibrations primarily capture high-frequency variations between the left and right wheel tracks, which are influenced by localized imperfections, rather than providing a complete representation of the road profile.

The dynamic model parameters representing the vehicle's physical behavior were estimated using chassis vibration measurements collected while driving over a laser-scanned road profile. These measurements were synchronized with road profile data through a \gls{gnss} matching algorithm, which probably introduced significant mismatches. To minimize sensitivity to mismatched input/output data, only the amplitude spectra of the measured and estimated vibrations were used. Although some parameters could not be estimated and were instead approximated, the estimated parameters still led to reasonable \gls{iri} estimates, suggesting that the system identification process was somewhat successful.

Despite these challenges, the estimation results could be improved with an enhanced system identification procedure. The current linear models do not account for the nonlinear behavior of vehicle suspensions, and more advanced models could offer better performance.

Although the \gls{iri} estimation method presented in this study performs well, there is room for improvement. For example, smoothing the state vector estimate in the \glsxtrlong{kf} by using $\hat{\bm{u}}[k|k+N]$ instead of $\hat{\bm{u}}[k|k+1]$ might enhance performance. Furthermore, experimenting with an adaptive process noise covariance $\bm{Q}$, which adjusts based on estimated \gls{iri} values, could further refine the accuracy of the estimates.

\bibliography{main}

\end{document}